\input harvmac
\input epsf
\noblackbox



\newcount\figno
\figno=0
\def\IZ{\relax\ifmmode\mathchoice
{\hbox{\cmss Z\kern-.4em Z}}{\hbox{\cmss Z\kern-.4em Z}}
{\lower.9pt\hbox{\cmsss Z\kern-.4em Z}}
{\lower1.2pt\hbox{\cmsss Z\kern-.4em Z}}\else{\cmss Z\kern-.4em
Z}\fi}
\def\IR{\relax{\rm I\kern-.18em R}}
\def\fig#1#2#3{
\par\begingroup\parindent=0pt\leftskip=1cm\rightskip=1cm\parindent=0pt
\baselineskip=11pt
\global\advance\figno by 1
\midinsert
\epsfxsize=#3
\centerline{\epsfbox{#2}}
\vskip 12pt
\centerline{{\bf Figure \the\figno} #1}\par
\endinsert\endgroup\par}
\def\figlabel#1{\xdef#1{\the\figno}}

\def\pmb#1{\setbox0=\hbox{#1}%
\kern-.025em\copy0\kern-\wd0
\kern.05em\copy0\kern-\wd0
\kern-.025em\raise.0433em\box0 }
\font\cmss=cmss10
\font\cmsss=cmss10 at 7pt
\def\half{{1\over 2}}
\def\rlx{\relax\leavevmode}
\def\Cop{\relax\,\hbox{$\kern-.3em{\rm C}$}}
\def\Rop{\relax{\rm I\kern-.18em R}}
\def\Nop{\relax{\rm I\kern-.18em N}}
\def\Pop{\relax{\rm I\kern-.18em P}}

\def\Zop{\rlx\leavevmode\ifmmode\mathchoice{\hbox{\cmss Z\kern-.4em Z}}
{\hbox{\cmss Z\kern-.4em Z}}{\lower.9pt\hbox{\cmsss Z\kern-.36em Z}}
{\lower1.2pt\hbox{\cmsss Z\kern-.36em Z}}\else{\cmss Z\kern-.4em
Z}\fi}


\def\ie{{\it i.e.}}
\def\eg{{\it e.g.}}

\lref\GiveonTQ{
A.~Giveon and D.~Kutasov,
JHEP {\bf 0001}, 023 (2000)
[hep-th/9911039].
}
\lref\WeinbergVP{
E.~J.~Weinberg and A.~Wu,
``Understanding Complex Perturbative Effective Potentials'',
Phys.\ Rev.\ D {\bf 36}, 2474 (1987).
}
\lref\GuthYA{
A.~H.~Guth and S.~Pi,
``The Quantum Mechanics Of The Scalar Field In The New Inflationary
Universe'',
Phys.\ Rev.\ D {\bf 32}, 1899 (1985).
}

\lref\MarcusVS{
N.~Marcus,
``Unitarity And Regularized Divergences In String Amplitudes'',
Phys.\ Lett.\ B {\bf 219}, 265 (1989).
}
\lref\FischlerTB{
W.~Fischler and L.~Susskind,
``Dilaton Tadpoles, String Condensates And Scale Invariance. 2'',
Phys.\ Lett.\ B {\bf 173}, 262 (1986).
}

\lref\SenMG{
A.~Sen,
``Non-BPS states and branes in string theory,''
hep-th/9904207.
}
\lref\MaldacenaRE{
J.~Maldacena,
``The large N limit of superconformal field theories and supergravity,''
Adv.\ Theor.\ Math.\ Phys.\  {\bf 2}, 231 (1998)
[Int.\ J.\ Theor.\ Phys.\  {\bf 38}, 1113 (1998)]
[hep-th/9711200].
}

\lref\WittenQJ{
E.~Witten,
``Anti-de Sitter space and holography'',
Adv.\ Theor.\ Math.\ Phys.\  {\bf 2}, 253 (1998)
[hep-th/9802150].
}

\lref\GubserBC{
S.~S.~Gubser, I.~R.~Klebanov and A.~M.~Polyakov,
``Gauge theory correlators from non-critical string theory,''
Phys.\ Lett.\ B {\bf 428}, 105 (1998)
[hep-th/9802109].
}

\lref\HarveyNA{
J.~A.~Harvey, D.~Kutasov and E.~J.~Martinec,
``On the relevance of tachyons'',
[hep-th/0003101].
}

\lref\AntoniadisAA{
I.~Antoniadis, C.~Bachas, J.~R.~Ellis and D.~V.~Nanopoulos,
``Cosmological String Theories And Discrete Inflation,''
Phys.\ Lett.\ B {\bf 211}, 393 (1988).
}

\lref\AntoniadisVI{
I.~Antoniadis, C.~Bachas, J.~R.~Ellis and D.~V.~Nanopoulos,
``An Expanding Universe In String Theory,''
Nucl.\ Phys.\ B {\bf 328}, 117 (1989).
}

\lref\AntoniadisUU{
I.~Antoniadis, C.~Bachas, J.~R.~Ellis and D.~V.~Nanopoulos,
``Comments On Cosmological String Solutions,''
Phys.\ Lett.\ B {\bf 257}, 278 (1991).
}

\lref\SeibergBJ{
N.~Seiberg and S.~H.~Shenker,
``A Note on background (in)dependence,''
Phys.\ Rev.\ D {\bf 45}, 4581 (1992)
[hep-th/9201017].
}

\lref\BalasubramanianDE{
V.~Balasubramanian, P.~Kraus, A.~E.~Lawrence and S.~P.~Trivedi,
``Holographic probes of anti-de Sitter space-times,''
Phys.\ Rev.\ D {\bf 59}, 104021 (1999)
[hep-th/9808017].
}

\lref\BalasubramanianSN{
V.~Balasubramanian, P.~Kraus and A.~E.~Lawrence,
``Bulk vs. boundary dynamics in anti-de Sitter spacetime,''
Phys.\ Rev.\ D {\bf 59}, 046003 (1999)
[hep-th/9805171].
}

\lref\BalasubramanianRY{
V.~Balasubramanian, S.~F.~Hassan, E.~Keski-Vakkuri and A.~Naqvi,
``A space-time orbifold: A toy model for a cosmological singularity,''
hep-th/0202187.
}

\lref\JackMK{
I.~Jack, D.~R.~Jones and J.~Panvel,
``Exact bosonic and supersymmetric string black hole solutions,''
Nucl.\ Phys.\ B {\bf 393}, 95 (1993)
[hep-th/9201039].
}

\lref\WittenYR{
E.~Witten,
``On string theory and black holes,''
Phys.\ Rev.\ D {\bf 44}, 314 (1991).
}

\lref\DijkgraafBA{
R.~Dijkgraaf, H.~Verlinde and E.~Verlinde,
``String propagation in a black hole geometry,''
Nucl.\ Phys.\ B {\bf 371}, 269 (1992).
}

\lref\BreitenlohnerBM{
P.~Breitenlohner and D.~Z.~Freedman,
``Positive Energy In Anti-De Sitter Backgrounds And Gauged Extended
Supergravity,''
Phys.\ Lett.\ B {\bf 115}, 197 (1982).
}

\lref\StromingerPN{
A.~Strominger,
``The ds/CFT correspondence,''
JHEP {\bf 0110}, 034 (2001)
[hep-th/0106113].
}

\lref\MaldacenaRE{
J.~Maldacena,
``The large $N$ limit of superconformal field theories and supergravity,''
Adv.\ Theor.\ Math.\ Phys.\  {\bf 2}, 231 (1998)
[Int.\ J.\ Theor.\ Phys.\  {\bf 38}, 1113 (1998)]
[hep-th/9711200].
}

\lref\AharonyUB{
O.~Aharony, M.~Berkooz, D.~Kutasov and N.~Seiberg,
``Linear dilatons, NS5-branes and holography,''
JHEP {\bf 9810}, 004 (1998)
[hep-th/9808149].
}

\lref\GasperiniHU{
M.~Gasperini and G.~Veneziano,
``Inflation, deflation, and frame independence in string cosmology,''
Mod.\ Phys.\ Lett.\ A {\bf 8}, 3701 (1993)
[hep-th/9309023].
}

\lref\VenezianoPZ{
G.~Veneziano,
``String cosmology: The pre-big bang scenario,''
hep-th/0002094.
}

\lref\BanksYP{
T.~Banks and W.~Fischler,
``M-theory observables for cosmological space-times,''
hep-th/0102077.
}

\lref\Gradshteyn{I. S. Gradshteyn and I. M. Ryzhik, {\it Table of
Integrals, Series and Products}, Academic Press, (1994).}.

\lref\KounnasWC{
C.~Kounnas and D.~L\"ust,
``Cosmological string backgrounds from gauged WZW models,''
Phys.\ Lett.\ B {\bf 289}, 56 (1992)
[hep-th/9205046].
}

\lref\TolleyCV{
A.~J.~Tolley and N.~Turok,
``Quantum fields in a big crunch / big bang spacetime,''
hep-th/0204091.
}

\lref\NekrasovKF{
N.~A.~Nekrasov,
``Milne universe, tachyons, and quantum group,''
hep-th/0203112.
}

\lref\LiuFT{
H.~Liu, G.~Moore and N.~Seiberg,
``Strings in a Time-Dependent Orbifold,''
hep-th/0204168.
}

\lref\SeibergHR{
N.~Seiberg,
``From big crunch to big bang - is it possible?,''
hep-th/0201039.
}

\lref\KhouryBZ{
J.~Khoury, B.~A.~Ovrut, N.~Seiberg, P.~J.~Steinhardt and N.~Turok,
``From big crunch to big bang,''
Phys.\ Rev.\ D {\bf 65}, 086007 (2002)
[hep-th/0108187].
}

\lref\ElitzurRT{
S.~Elitzur, A.~Giveon, D.~Kutasov and E.~Rabinovici,
``From Big Bang to Big Crunch and Beyond,''
hep-th/0204189.
}

\lref\Vilenkin{
N.~J.~Vilenkin, ``Special Functions and the Theory of Group
Representations'',
AMS, 1968.}

\lref\BrandenbergerAJ{
R.~H.~Brandenberger and C.~Vafa,
``Superstrings In The Early Universe,''
Nucl.\ Phys.\ B {\bf 316}, 391 (1989).
}

\lref\DiFrancescoUD{
P.~Di Francesco and D.~Kutasov,
``World sheet and space-time physics in two-dimensional (Super)string
theory,''
Nucl.\ Phys.\ B {\bf 375}, 119 (1992)
[hep-th/9109005].
}
\lref\KutasovPV{
D.~Kutasov,
``Some properties of (non)critical strings,''
hep-th/9110041.
}

\lref\CornalbaNV{
L.~Cornalba, M.~S.~Costa and C.~Kounnas,
``A resolution of the cosmological singularity with orientifolds,''
hep-th/0204261.
}

\lref\CornalbaFI{
L.~Cornalba and M.~S.~Costa,
``A New Cosmological Scenario in String Theory,''
hep-th/0203031.
}

\lref\TeschnerFT{
J.~Teschner,
``On structure constants and fusion rules in the SL(2,C)/SU(2) WZNW  model,''
Nucl.\ Phys.\ B {\bf 546}, 390 (1999)
[hep-th/9712256].
}

\lref\TseytlinXK{
A.~A.~Tseytlin and C.~Vafa,
``Elements of string cosmology,''
Nucl.\ Phys.\ B {\bf 372}, 443 (1992)
[hep-th/9109048].
}

\lref\LustVD{
D.~L\"ust,
``Cosmological string backgrounds,''
hep-th/9303175.
}

\lref\MuellerIN{
M.~Mueller,
``Rolling Radii And A Time Dependent Dilaton,''
Nucl.\ Phys.\ B {\bf 337}, 37 (1990).
}

\lref\PolchinskiRQ{
J.~Polchinski,
``String Theory. Vol. 1: An Introduction To The Bosonic String,''
{\it  Cambridge, UK: Univ. Pr. (1998) 402 p}.
}

\lref\ColeyUH{
A.~A.~Coley, ``Dynamical systems in cosmology,''
gr-qc/9910074.
}

\lref\Abbott{
L.~F.~Abbott and S.~Deser,
``Stability Of Gravity With A Cosmological Constant,''
Nucl.\ Phys.\ B {\bf 195}, 76 (1982).}

\lref\BirrellIX{
N.~D.~Birrell and P.~C.~Davies,
``Quantum Fields In Curved Space,''
{\it  Cambridge, Uk: Univ. Pr. ( 1982) 340p}.
}

\lref\MezincescuEV{
L.~Mezincescu and P.~K.~Townsend,
``Stability At A Local Maximum In Higher Dimensional Anti-De Sitter Space
And Applications To Supergravity,''
Annals Phys.\  {\bf 160}, 406 (1985).
}

\lref\BreitenlohnerBM{
P.~Breitenlohner and D.~Z.~Freedman,
``Positive Energy In Anti-De Sitter Backgrounds And Gauged Extended
Supergravity,''
Phys.\ Lett.\ B {\bf 115}, 197 (1982).
}

\lref\BreitenlohnerJF{
P.~Breitenlohner and D.~Z.~Freedman,
``Stability In Gauged Extended Supergravity,''
Annals Phys.\  {\bf 144}, 249 (1982).
}

\lref\HassanMQ{
S.~F.~Hassan and A.~Sen,
``Twisting classical solutions in heterotic string theory,''
Nucl.\ Phys.\ B {\bf 375}, 103 (1992)
[hep-th/9109038].
}

\lref\SenZI{
A.~Sen,
``O(d) x O(d) symmetry of the space of cosmological solutions in string
theory, scale factor duality and two-dimensional black holes,''
Phys.\ Lett.\ B {\bf 271}, 295 (1991).
}

\lref\GasperiniAK{
M.~Gasperini and G.~Veneziano,
``O(d,d) covariant string cosmology,''
Phys.\ Lett.\ B {\bf 277}, 256 (1992)
[hep-th/9112044].
}

\lref\MeissnerZJ{
K.~A.~Meissner and G.~Veneziano,
``Symmetries of cosmological superstring vacua,''
Phys.\ Lett.\ B {\bf 267}, 33 (1991).
}

\lref\MeissnerGE{
K.~A.~Meissner and G.~Veneziano,
``Manifestly O(d,d) invariant approach to space-time dependent string
vacua,''
Mod.\ Phys.\ Lett.\ A {\bf 6}, 3397 (1991)
[hep-th/9110004].
}

\lref\SmithUP{
E.~Smith and J.~Polchinski,
``Duality survives time dependence,''
Phys.\ Lett.\ B {\bf 263}, 59 (1991).
}

\lref\SilversteinXN{
E.~Silverstein,
``(A)dS backgrounds from asymmetric orientifolds,''
hep-th/0106209.
}

\lref\AharonyCX{
O.~Aharony, M.~Fabinger, G.~Horowitz and E.~Silverstein,
``Clean time-dependent string backgrounds from bubble baths,''
hep-th/0204158.
}

\lref\TseytlinWR{
A.~A.~Tseytlin,
``Duality and dilaton,''
Mod.\ Phys.\ Lett.\ A {\bf 6}, 1721 (1991).
}

\lref\TseytlinHT{
A.~A.~Tseytlin,
``On the form of the black hole solution in D = 2 theory,''
Phys.\ Lett.\ B {\bf 268}, 175 (1991).
}

\lref\SimonMA{
J.~Simon,
``The geometry of null rotation identifications,''
hep-th/0203201.
}

\lref\BanksFE{
T.~Banks,
``Cosmological breaking of supersymmetry or little Lambda goes back to  the future. II,''
hep-th/0007146.
}

\lref\SusskindRI{
L.~Susskind,
``Twenty years of debate with Stephen,''
hep-th/0204027.
}

\lref\SpradlinPW{
M.~Spradlin, A.~Strominger and A.~Volovich,
``Les Houches lectures on de Sitter space,''
hep-th/0110007.
}

\lref\BoussoNF{
R.~Bousso,
``Positive vacuum energy and the N-bound,''
JHEP {\bf 0011}, 038 (2000)
[hep-th/0010252].
}

\lref\MaldacenaMW{
J.~M.~Maldacena and C.~Nunez,
``Supergravity description of field theories on curved manifolds and a no  go theorem,''
Int.\ J.\ Mod.\ Phys.\ A {\bf 16}, 822 (2001)
[hep-th/0007018].
}
\Title{\vbox{
\hbox{hep--th/0205101}
\hbox{EFI-02-77}}}
{\vbox{\centerline{String Propagation in the Presence}
\vskip 10pt
\centerline{of Cosmological Singularities}}}
\centerline{Ben Craps\footnote{$^\star$}{{\tt
craps,kutasov,rajesh@theory.uchicago.edu}}$^{,a}$,
David Kutasov$^{\star,a,b}$ and
Govindan Rajesh$^{\star,a}$
}
\bigskip
\centerline{\it $^a$Enrico Fermi Institute, University of Chicago,
5640 S. Ellis Av., Chicago, IL 60637, USA}
\medskip
\centerline{\it $^b$Department of Physics, University of Chicago,
5640 S. Ellis Av., Chicago, IL 60637, USA}
\bigskip
\bigskip
\noindent
We study string propagation in a spacetime with
positive cosmological constant, which includes
a circle whose radius approaches a finite value as
$|t|\to\infty$, and goes to zero at $t=0$. Near this
cosmological singularity, the spacetime looks like
$\IR^{1,1}/\IZ$. In string theory, this spacetime
must be extended by including four additional regions,
two of which are compact. The other two introduce new
asymptotic regions, corresponding to early
and late times, respectively. States of quantum
fields in this spacetime are defined in the tensor product
of the two Hilbert spaces corresponding to the early time
asymptotic regions, and the S-matrix describes the evolution
of such states to states in the tensor product of the two late
time asymptotic regions. We show that string theory provides a
unique continuation of wavefunctions past the cosmological
singularities, and allows one to compute the S-matrix. The
incoming vacuum evolves into an outgoing state with particles.
We also discuss instabilities of asymptotically timelike linear
dilaton spacetimes, and the question of holography in such
spaces. Finally, we briefly comment on the relation of our results
to recent discussions of de Sitter space.

\bigskip

\Date{5/02}

\newsec{Introduction}

The purpose of this paper is to study some time-dependent solutions in
string theory with a positive cosmological constant. At early and late
times $(|t|\to\infty)$, the spacetimes we will consider asymptote to
linear dilaton solutions, with string frame metric and dilaton
\eqn\timedil{\eqalign{
ds^2=&-dt^2+dx^idx^i~;\cr
e^{\Phi}=&e^{-Q|t|}~,\cr
}}
where $x^i$, $i=1,\cdots,d$ are spatial coordinates\foot{One can also
replace the flat space labeled by $x_i$ by a more general background,
corresponding to a CFT ${\cal M}$ with central charge $d$, which is
not necessarily integer.}, and the linear dilaton slope $Q$ is related
to the dimension of space $d$ in a way described below. Thus, the theory
becomes arbitrarily weakly coupled in the far past and far future.

The finite time behavior is non-trivial, and generically
these solutions contain cosmological singularities. One
of the issues of interest is the physics associated with
these singularities. In particular, it is in general not
clear how to define observables in a cosmological spacetime:
should one specify initial conditions near a singularity
\BanksYP, or include a ``pre-big bang'' period \VenezianoPZ\
and specify initial conditions in an ``asymptotic past trivial''
regime? We will see below that string theory in the backgrounds
\timedil\ seems to favor the latter. 

In fact, in the example studied in this 
paper, we will find that the global spacetime contains two incoming and two
outgoing regions, connected to each other in a non-trivial way. String theory
provides a prescription for continuing wavefunctions through the singularities
separating the various regions. This allows one to study an S-matrix for string
propagation through the singularities, and thereby obtain information about
physics in the vicinity of the singularity. In particular, we will determine the
natural incoming and outgoing vacua, and compute the Bogolubov coefficients
relating them. We will find that the incoming vacuum evolves to a state with
particles.

Another interesting issue concerns the stability of the solutions \timedil.
In spacetimes with positive cosmological constant, SUSY is usually broken
and one must check for the existence of growing modes. We will show that
asymptotically timelike linear dilaton spacetimes in general contain 
non-negative mass squared modes whose wavefunctions grow exponentially with 
time at early and late times. The
contribution of these modes to one loop (torus) amplitudes is infrared
divergent, like that of a tachyon in flat spacetime with constant dilaton.
In particular, the graviton is tachyonic in this case.

One of the motivations for this work is the question of
holography in gravity with
positive cosmological constant. In \StromingerPN\ it
was proposed that gravity in de Sitter spacetime
is dual to a Euclidean CFT living on the boundary at
$|t|\to\infty$. This duality is modeled after the
AdS/CFT correspondence for negative cosmological
constant. In particular, on-shell wavefunctions in de
Sitter space with specified behavior near the boundary
are expected to be dual to off-shell operators in the
CFT, and S-matrix elements in de Sitter space should
correspond to Green functions in the boundary theory. Many
issues concerning this duality remain unclear. For example,
one finds that while low mass fields in the bulk are dual
to operators with real scaling dimensions in the boundary
CFT, above a certain critical value of the mass the scaling
dimensions become complex. Also, it is not clear whether the
duality requires the presence of both past and future boundaries,
or whether one of them is sufficient (this is related to the
question of observables mentioned above).

Some of these issues can be studied in asymptotically
timelike linear dilaton spacetimes, which share many
properties with (global) de Sitter space. Both contain
spacelike boundaries\foot{In the linear dilaton case,
the boundary is the weak coupling region, $|t|\to\infty$,
as in \AharonyUB\ in the spacelike case.}
at early and late times, and some
properties of solutions of the wave equation near the
boundary are very similar. As in the case of de Sitter
space, in asymptotically timelike linear dilaton vacua
there is a positive value of the mass squared at which
the qualitative behavior of the solutions changes. The main
advantage of timelike linear dilaton solutions is that they
are easy to embed in string theory, which allows these issues
to be studied in a relatively controlled setting.

Another reason to expect that timelike linear dilaton
solutions should be useful for studying holography in
de Sitter space is the analogy to the case of negative
cosmological constant. In that case, it is known that
AdS and spacelike linear dilaton vacua have a
holographic description in terms of CFT \MaldacenaRE\
and Little String Theory \AharonyUB, respectively. Many
of the issues related to holography in AdS and spacelike
linear dilaton spacetimes are
similar. For example, in both cases non-normalizable
on-shell bulk wavefunctions correspond to operators in
the boundary theory; the Breitenlohner-Freedman bound of
AdS space \BreitenlohnerBM\
has a precise analogue in spacelike linear dilaton theories;
issues related to stability are similar in the two cases.
Thus, it is natural to expect that if the AdS/CFT
correspondence has an analogue in de Sitter space, something
similar should happen for timelike linear dilaton spacetimes.
Of course, one does not expect timelike linear dilaton
spacetimes to be dual to CFT's, just as in the case of
negative cosmological constant, but the more qualitative
issues mentioned above should have counterparts in the linear
dilaton case.

\lref\WittenKN{
E.~Witten,
``Quantum gravity in de Sitter space,''
hep-th/0106109.
}

With this in mind, we discuss below the
implications of our results for holography
in de Sitter and timelike linear dilaton
spacetimes. We find that the physical picture
is rather different for positive and negative
cosmological constant. The
observables\foot{or, in the de Sitter case,
meta-observables \WittenKN.}
in de Sitter space and timelike linear dilaton
backgrounds are more similar to the scattering
states of flat space string theory than to the
non-normalizable observables familiar from AdS.
Correlation functions of these observables
compute S-matrix elements of the scattering
states. Wavefunctions which give rise to real
scaling dimensions in the boundary CFT correspond
to growing modes and lead
to infrared divergences in loop amplitudes.

The plan of the paper is as follows.
In section 2 we discuss some classical solutions
in gravity with positive cosmological constant.
At large $|t|$ these solutions asymptote to
\timedil. They exhibit a cosmological singularity
at $t=0$, near which they look like generalized Kasner
solutions. We focus on a particular case, the generalized
Milne universe, in which the geometry includes a circle
whose radius shrinks from a finite value at $t=-\infty$
to zero at $t=0$, and then increases again to the same
finite value at $t=+\infty$. Near the big crunch/big bang
singularity at $t=0$ the geometry looks like $\IR^{1,1}/\IZ$
(and thus has regions with closed timelike curves). We point
out that it is natural to continue the spacetime to one which
has four asymptotic regions, two corresponding to
early times, and two to late times (see figure 2).

In section 3 we embed the generalized Milne universe
into string theory. We discuss the structure of the Hilbert
space of asymptotic states in each of the two incoming
(or, alternatively, outgoing) regions in figure 2, and
show that the full spacetime can be viewed as a certain
coset of $SL(2,\IR)$.

In section 4 we study perturbations of the generalized
Milne universe, associated with a minimally coupled
scalar field of mass $m$, using the $SL(2,\IR)$ description
to continue wavefunctions through the singularities.
We show that the Hilbert space
of in-states is the direct product of the Hilbert spaces
corresponding to the two in-regions in figure 2, and
similarly for the Hilbert space of out-states. We compute
the Bogolubov coefficients relating the natural creation
and annihilation operators at early times to those at late
times. The in-vacuum corresponds to a state
with particles in the out-Hilbert space.

In section 5 we discuss the growing modes which are
typically present in asymptotically timelike linear
dilaton spacetimes \timedil. We show that these modes,
which have $m^2<Q^2$, give rise to infrared
divergences in one loop amplitudes.

In section 6 we comment on the relation of our results to
gravity in asymptotically de Sitter space, which shares
many of the properties discussed in the asymptotically
timelike linear dilaton case. In particular, we point out
that fields with mass smaller than the Hubble mass have
properties similar to those of fields with $m^2<Q^2$
in the timelike linear dilaton case.

In section 7 we summarize our results and comment on them.

\newsec{Some classical solutions with positive cosmological constant}

We start with a discussion of some solutions of
gravity coupled to a dilaton, with a positive
cosmological constant. In the applications discussed
in the following sections, the gravity approximation
is in general invalid, because the gradient of the
dilaton is typically of order one in string units,
and due to the presence of (timelike and spacelike)
singularities. Nevertheless, the gravity analysis provides
a useful qualitative guide to the structure of spacetime.
In some cases, the gravity solution does not receive
stringy corrections and is exact, even in the presence
of singularities. The behavior of fields near singularities
is in general ambiguous in general relativity, and will be dealt
with in the next sections, using ideas from string theory.

In $d+1$ dimensional spacetime, string theory gives rise
to a classical low-energy effective action of the form
\eqn\action{
S_g = {1\over 2\kappa^2} \int{d^{d+1}x\sqrt{-g}
e^{-2\Phi}\bigl[R + 4 g^{\mu\nu} \partial_\mu
\Phi \partial_\nu\Phi - 2\Lambda]}~,
}
where $\kappa^2=8\pi G_N$. The fields $\Phi$ and $g$ are
the $d+1$ dimensional dilaton and metric,
respectively. The other massless and massive
modes of the string will be set to zero for now;
we will discuss them in the next sections, when
we study perturbations.\foot{Ramond-Ramond fields
have recently been included in \refs{\SilversteinXN},
with the aim of constructing de Sitter solutions in
supercritical string theory.}
A non-zero tree level cosmological constant
$\Lambda$ can be obtained, for example, by
considering non-critical strings (see \eg\
\PolchinskiRQ, p. 114), but for
now we will treat it as a free parameter.

We look for solutions for which the dilaton
$\Phi$ and the diagonal components of the string
frame metric $g$ depend on time but not on
space:\foot{More general solutions can be obtained
by applying O($d,d$) transformations
\refs{\MeissnerZJ,\MeissnerGE,\SenZI,\HassanMQ,\GasperiniAK}.}
\eqn\kasmet{
ds^2 = - dt^2 + \sum_{i=1}^{d}e^{2\alpha_i(t)} dx_i^2~.
}
If $x_i$ is compact, it is convenient to take it
to live on a circle of radius $l_s=\sqrt{\alpha'}$.
$R_i(t)=l_s\exp(\alpha_i(t))$ is then the dynamical
radius of the $i$'th dimension. In the rest of this
paper we will set $l_s=1$.

The equations of motion of $\alpha_i(t)$, $\Phi(t)$
which follow from the action \action\ can be
derived from the mini-superspace action
\eqn\mini{S_g=-{1\over 2\kappa^2}
\int dt \sqrt{-g_{00}}{e^{-\phi}\left(
\sum_{i=1}^{d}g^{00}\dot\alpha_i^2 -
g^{00}\dot\phi^2 +
2\Lambda\right)}~,
}
where $\phi$ is defined by
\eqn\phidef{\phi=2\Phi - \sum_{i=1}^{d}\alpha_i~.}
One can think of $\exp(\phi/2)$ as the effective string
coupling of the lower-dimensional theory obtained by
averaging over $x_i$. The equations of motion which
follow from the action \mini\ are~\refs{\MuellerIN, \LustVD, \ColeyUH}
\eqn\kaseom{
\eqalign{
\ddot{\alpha_i} - \dot{\alpha_i} \dot{\phi} &= 0~; \cr
2 \ddot{\phi} - \dot{\phi}^2 -  \sum_{i=1}^{d}\dot{\alpha_i}^2
+ 2 \Lambda &= 0~;\cr
\sum_{i=1}^{d}\dot{\alpha_i}^2 - \dot{\phi}^2 + 2 \Lambda &= 0 ~.\cr}}
The last equation is obtained by varying with respect to $g_{00}$,
before fixing the gauge $g_{00}=-1$. Equations \kaseom\ can be
reduced to:
\eqn\kassoln{
\eqalign{
\dot\alpha_i &= c_i e^{\phi}~;\cr
\ddot\phi &= \dot\phi^2 - 2 \Lambda = \sum_{i=1}^{d} \dot\alpha_i^2~,\cr}}
where $c_i$ are arbitrary (real) integration constants.

To understand the qualitative structure of the solutions
of \kassoln, it is useful to note that one can think
of the dilaton $\phi$ as the position (on an infinite line)
of a non-relativistic particle of unit mass. Substituting
the first line of \kassoln\ into the second then gives the
equation of motion of the particle in the potential
\eqn\vvpphhii{V(\phi) = - {1\over 2}e^{2\phi} \sum_{i=1}^d c_i^2 ~,}
with $\Lambda$ playing the role of the total energy of the particle.
We next discuss some solutions of these equations that are going to
be of interest below.

\subsec{Timelike linear dilaton}

A simple solution of \kassoln\ is obtained by setting all the
integration constants $c_i$ to zero. In this case, the potential
\vvpphhii\ vanishes, and the auxiliary particle moves with constant
velocity determined by its total energy $\Lambda$. All $\alpha_i$ are
constant, and the string-frame metric is flat. The ``shifted'' dilaton
$\phi$ is simply twice the true dilaton (see \phidef), and~\kassoln\ can
be solved to give
\eqn\tldsol{ds^2=-dt^2+dx_i^2,\;\;\Phi = \pm \sqrt{\Lambda\over 2}\; t~.}
As we review in section 3, the solution \tldsol\ corresponds to
a simple exact CFT (for any $\Lambda$), but unfortunately it is
singular. Indeed,
this solution has the property that the string coupling $g_s=\exp(\Phi)$
diverges either at large positive time (for the $+$ sign in \tldsol)
or at large negative time (for the $-$ sign). Thus, the physics of
this model is not perturbative -- correlation functions are typically
pushed into the strong coupling region (see \eg\
\refs{\DiFrancescoUD, \KutasovPV} for discussions).

For $t\not=0$, one can consider the solution
\eqn\tldabs{\alpha_i={\rm const},\;\;
\Phi = - \sqrt{\Lambda\over 2}\; |t|~,}
for which the string coupling does not necessarily become
large anywhere. However, \tldabs\ does not solve the sourceless
equations of motion at $t=0$, and is thus incomplete.
In subsection {\it 2.2},
we will construct solutions which look like \tldabs\ for large
$|t|$ but resolve the apparent non-analyticity near $t=0$.

If the dimension of spacetime is larger than two ($d>1$),
one can pass to the Einstein frame, which is useful for some
purposes. Performing the standard Weyl transformation
\PolchinskiRQ\ on the string frame metric \tldsol,
\eqn\weyltr{g_{\mu\nu,E}=e^{-{4\Phi\over d-1}}g_{\mu\nu}~,}
one finds the line element
\eqn\einsteinmetric{
ds^2_E=e^{4Qt\over d-1}\left(
-dt^2+dx_i^2\right)~,
}
where $Q=\mp\sqrt{\Lambda/2}$. We will take $Q$ to be positive,
\eqn\Qdef{Q=\sqrt{\Lambda\over2}~,}
so that the string coupling
\eqn\ggss{g_s=e^{-Qt}}
goes to zero at late times. Defining a new time coordinate
\eqn\tX{
\tau={d-1\over 2Q}\,e^{2Qt\over d-1}~,
}
\einsteinmetric\ becomes
\eqn\einsteinmetricbis{
ds^2_E=-d\tau^2+\left({2Q\over d-1}\right)^2 \tau^2dx_i^2~,
}
an FRW metric with a linearly growing scale factor \AntoniadisAA
\eqn\at{a(\tau)={2Q\over d-1} \tau~.}
The Ricci scalar of an FRW metric with zero spatial curvature is given by
\eqn\ricci{
{\cal R}=2d{\ddot a\over a}+(d^2-d)\left({\dot a\over a}\right)^2~.
}
For the metric \einsteinmetricbis\ we find
\eqn\riccitld{
{\cal R}={d^2-d\over \tau^2}~,
}
so that there is a curvature singularity at $\tau=0$.
Since $\tau=0$ corresponds to $t=-\infty$ (see \tX),
this singularity is not surprising. The string coupling
\ggss\ diverges as $\tau\to 0$, and one does not expect
the solution in that region to be reliable.

We will mostly focus below on the string frame metric
$g_{\mu\nu}$ \action, since this is the metric felt
by fundamental strings, which will be our main focus.
The Einstein frame metric is useful
for comparing to solutions of general relativity, and
for studying other probes in the theory.

\subsec{Generalized Kasner}

Since the solution with all $c_i$ \kassoln\ set to zero is singular,
we next turn to solutions in which some or all of the $c_i$ are
non-zero. In fact, it is not difficult to solve
the equations of motion \kassoln\ with generic $c_i$.
The solution has been described in~\LustVD, but in order to keep
the discussion self-contained, we will review it here.
We can first solve for $\phi$ by using the auxiliary description in terms
of a particle rolling in the potential \vvpphhii, and then substitute
the solution in \kassoln\ to find the scale factors $\alpha_i$. We
furthermore impose the boundary condition that at early and late times
the solution should approach the weakly coupled timelike linear dilaton
one: $\alpha_i={\rm const}$, $\phi=-2Q|t|$. We find
\eqn\phidotsol{
\dot\phi = - 2Q\; {\rm coth} (2Q t)~,}
so that
\eqn\phisol{
\phi = -{1\over 2}{\rm log}({\rm sinh}^2 (2Q t)) + C~.}
The effective string coupling $\exp\phi/2$ becomes large near $t=0$,
and one may expect a cosmological singularity at $t=0$. Note also
that the solution is symmetric under time reversal, $t\to -t$. Due to
the singularity at $t=0$, we are really solving the equations of motion
separately for positive and negative $t$. The question of matching
the positive and negative $t$ solutions across the singularity will
be addressed for a special case in subsections~{\it 2.3} and~{\it 3.2}.

The constant of integration $C$ in \phisol\ is in principle arbitrary,
but one natural way to fix it is the following. For $t\to 0$, the
``velocity of the particle'' $\dot\phi$ becomes very large and one
can neglect the total energy $\Lambda$, relative to the kinetic
energy $\half\dot\phi^2$, and the potential energy $V(\phi)$, \vvpphhii,
separately. Thus, for small $t$ one can neglect the cosmological
constant $\Lambda$ in the cosmological equations \kaseom. The resulting
equations have been studied in the context of pre-big bang
scenarios (see \eg\ \refs{\GasperiniHU, \VenezianoPZ}); they have a
well-known class of solutions corresponding to generalized Kasner or
homogeneous Bianchi I spacetimes. It is natural to require that in the
limit $t\to 0$, our solutions reduce to those of
\refs{\GasperiniHU, \VenezianoPZ}. This fixes $C$ in \phisol.
One finds
\eqn\kasnersoln{
\eqalign{
\phi &= -{1\over 2} {\rm log}\left[{1\over 4Q^2} \left(\sum_i{c_i^2}\right)
\;
{\rm sinh}^2 (2Qt)\right]~;\cr
\alpha_i &= {a_i\over2} \;
{\rm log}\Bigl[{1\over Q^2}
{\rm tanh}^2 (Qt)\Bigr]~,\cr
}}
where
\eqn\aaii{a_i\equiv {c_i\over\sqrt{\sum_j{c_j^2}}}~.}
Note that $a_i$ satisfy $\sum_i a_i^2=1$.

Another way of presenting the solution \kasnersoln\
is in terms of the higher
dimensional dilaton $\Phi$ \phidef,
\eqn\highphi{e^{2\Phi}={2Q\over\sqrt{\sum_j c_j^2\sinh^2(2Qt)}}
\left({1\over Q^2}\tanh^2 (Qt)\right)^{\sum_i \half a_i}
}
and the radii of the spatial dimensions,
\eqn\rrii{R_i(t)=\left({1\over Q^2}\tanh^2 (Qt)\right)^{\half a_i}~.
}
Depending on the sign of $a_i$, the $i$'th dimension is either
expanding or contracting with time. Positive $a_i$ corresponds to
pre-big bang contraction, while negative $a_i$ corresponds to
expansion (in the string frame).

\lref\GiveonSY{
A.~Giveon,
``Target space duality and stringy black holes,''
Mod.\ Phys.\ Lett.\ A {\bf 6}, 2843 (1991).
}

In string theory one expects
\eqn\tici{T_i:\;a_i\to -a_i}
for any $i$ to be a symmetry \refs{\TseytlinWR,\TseytlinXK}. This
transformation inverts the
radius \rrii\ $R_i(t)\to 1/R_i(t)$, and thus acts as
a time-dependent T-duality \refs{\SmithUP}. The fact that the lower
dimensional
dilaton $\phi$ \kasnersoln\ is invariant under the transformation
$T_i$ \tici\ is also consistent with this interpretation.
The symmetry \tici\ is analogous to the familiar T-duality
relating the Euclidean cigar and trumpet \refs{\GiveonSY,\DijkgraafBA},
with the radial direction of the cigar or trumpet replaced by
the time, $t$.

\subsec{Generalized Milne}

For generic $c_i$, the solution \highphi, \rrii\ has a cosmological
singularity at $t=0$. An interesting special case which is less
singular is obtained by setting
$c_1=a_1=1$ (and thus all the other $c_i=0$). In
this case, the geometry is that of flat $d-1$ dimensional space times
a two dimensional spacetime with metric~\refs{\KounnasWC, \LustVD}
\eqn\bhmet{ds^2 = - dt^2 +{1\over Q^2}
\tanh^2(Qt)\; dx^2}
and dilaton
\eqn\bhdil{\Phi(t)= -\log\cosh(Qt)~.}
Rescaling $t\rightarrow t/Q$, one finds
\eqn\bhsol{\eqalign{
ds^2 &= {1\over Q^2}\; (-dt^2 + {\rm tanh}^2t\;dx^2)~;\cr
\Phi &= - {\rm log}\cosh t~.\cr}}
Near $t=0$, \bhsol\ reduces to
\eqn\bhsolbis{\eqalign{
ds^2 &\sim {1\over Q^2}\; (-dt^2 + t^2\, dx^2)~;\cr
\Phi &\sim 0~.\cr}}
The structure of spacetime near $t=0$ depends on whether
the spatial coordinate $x$ is compact or not.

If $x$ is
non-compact, the spacetime \bhsolbis\ is completely smooth.
Indeed, the metric \bhsolbis\ is flat (see \ricci),
and the dilaton is constant. This
is the two dimensional Milne universe,
an unconventional parametrization
of two wedges in flat Minkowski
spacetime. To get the full Minkowski spacetime one must
add two Rindler wedges (obtained by
continuing $t$ to imaginary values in \bhsolbis).

\midinsert\bigskip{\vbox{{\epsfxsize=3in
        \nobreak
    \centerline{\epsfbox{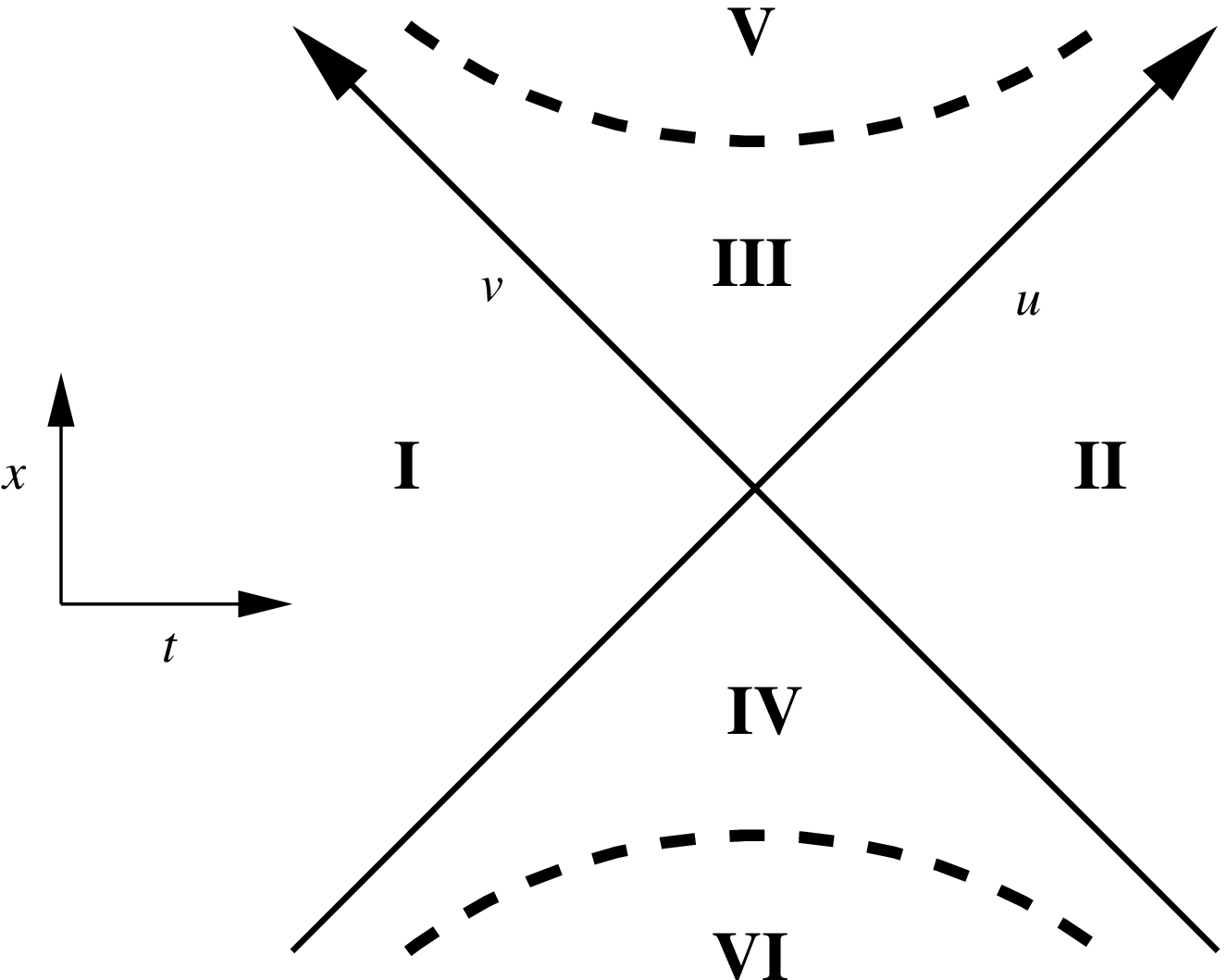}}
        \nobreak\bigskip
    {\raggedright\it \vbox{
{\bf Fig 1.}
{\it The generalized Milne universe.}}}}}}
\bigskip\endinsert

A similar continuation in the full geometry \bhsol\ is
achieved by changing coordinates to
\eqn\newcoor{
\eqalign{
u=\sinh t\, e^{-x}~;\cr
v=-\sinh t\, e^x~,
}
}
in terms of which the background \bhsol\ is given
by
\eqn\bhuv{
\eqalign{
ds^2&={1\over Q^2} {du\,dv\over1-uv}~;\cr
\Phi&=
-{1\over4}{\rm log}(1-uv)^2~.
}
}
The original $(t,x)$ coordinates cover the two wedges
$uv<0$ (regions $I$, $II$ in figure~1). The global spacetime
corresponds to $-\infty<u,v<\infty$. It contains a curvature
singularity and strong coupling region at $uv=1$. Following
the discussion in subsection {\it 2.1}, one might be worried
that the singularity might invalidate a weak coupling treatment.
We will see in section~4 that this does not seem to be the case.

\lref\HorowitzAP{
G.~T.~Horowitz and A.~R.~Steif,
``Singular String Solutions With Nonsingular Initial Data,''
Phys.\ Lett.\ B {\bf 258}, 91 (1991).
}

The geometry \bhsol, which describes regions $I$ and $II$ in
figure~1, is invariant under translations of $x$;
thus, it is natural to ask what happens when $x$ is periodically
identified, such that the early and late time geometry \bhsol\
is $\IR\times S^1$. The resulting spacetime is singular; the
spatial circle in region $I$ shrinks from a finite size at
large negative $t$ to zero size at $t=0$, and then expands
again for positive $t$ (in region $II$).
Thus, the identification $x\simeq x+2\pi R$ which is spacelike
at large $|t|$, becomes null as $t\to 0$, and can be thought of
as an identification by a boost. This gives rise to a
spacetime of the kind discussed in~\refs{\HorowitzAP,
\KhouryBZ,\SeibergHR, \CornalbaFI, \NekrasovKF, \SimonMA, \TolleyCV,
\LiuFT, \ElitzurRT, \CornalbaNV}.
In terms of the variables \newcoor, which cover the whole
spacetime of figure~1, the identification is
\eqn\uvid{(u,v)\to (u e^{-\tau}, v e^\tau)}
with $\tau=2\pi R$. Near $uv=0$, the manifold looks like
$\IR^{1,1}/\IZ$. There are points with $uv=0$ that are distinct but
cannot be separated by open sets -- the spacetime is non-Hausdorff.
For instance, any point with $uv=0$ is identified by \uvid\ with points
arbitrarily close to $u=v=0$.

For $uv<0$ (regions $I$, $II$ in figure~1),
the identification \uvid\ is spacelike (as we saw before),
while for $0<uv<1$ (regions $III$ and $IV$ in figure 1) it
is timelike. Thus, the latter region contains closed timelike
curves. For $uv>1$ (regions $V$ and $VI$),
the identification becomes spacelike again.
In fact, large positive $uv$ corresponds to another
pair of asymptotic early and late time regions.

The resulting spacetime is qualitatively of the form
depicted in figure 2. This figure is intended to give
a rough idea of the structure of asymptotic regions, but
is less precise in the structure near the singularities.
For example, the metric in region $VI$ is in fact
such that the radius of the circle is increasing
as one moves in towards the singularity, and the string
coupling grows as well. In drawing figure 2 we have
implicitly performed a T-duality locally on that region,
to bring it to a form like that of region $I$.

Due to the appearance of a cosmological singularity for
compact $x$, many aspects of QFT in the spacetime \bhsol,
\bhuv\ are unclear. In particular, it is not a priori
obvious whether one should restrict to the expanding,
post-big bang, part of the spacetime (region $II$ in
figures 1,2), or include the other regions indicated
in figures 1,2. In the latter case, one must understand
how to continue the wavefunctions across cosmological
singularities.

\midinsert\bigskip{\vbox{{\epsfxsize=3in
        \nobreak
    \centerline{\epsfbox{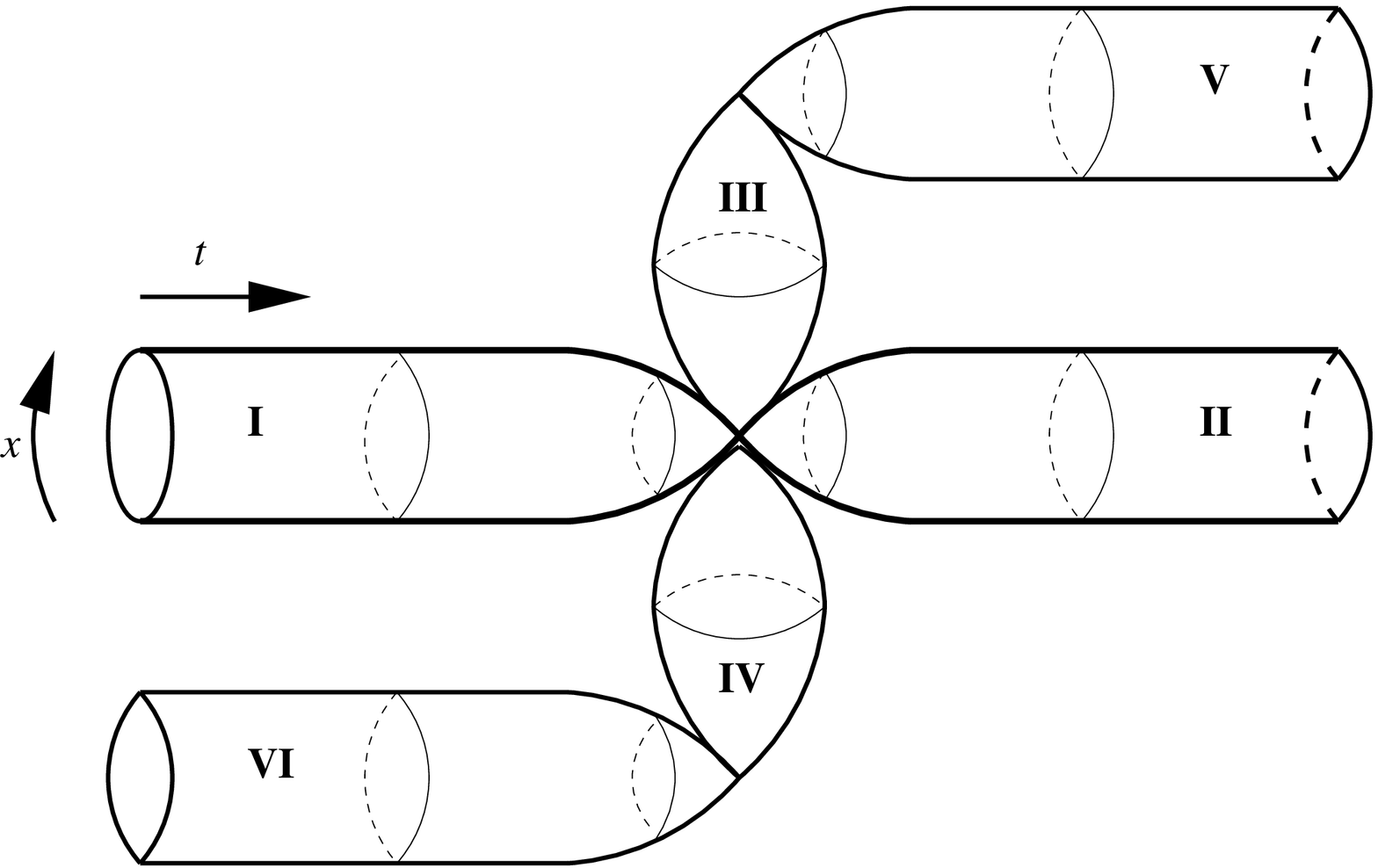}}
        \nobreak\bigskip
    {\raggedright\it \vbox{
{\bf Fig 2.}
{\it The periodically identified generalized Milne solution.}}}}}}
\bigskip\endinsert

The approach to this problem that we will take is to realize the
spacetime \bhsol, \bhuv\ as an exact classical solution in string
theory. As we will see in the next sections, string theory provides
a coherent picture of the physics in the full spacetime \bhuv.
It describes both the ``pre-big bang'' regime $t<0$ (region~I)
and the ``post-big bang'' regime $t>0$ (region~II), as well as the
other regions
in figures 1,2, and gives rise to well defined matching conditions
of wavefunctions across the singularities. It can thus be used
to evolve fields through the singularities, and study string
interactions in the spacetime \bhsol, \bhuv.

\lref\MyersFV{
R.~C.~Myers,
``New Dimensions For Old Strings,''
Phys.\ Lett.\ B {\bf 199}, 371 (1987).
}

\newsec{String theory}

In this section we describe some features of
string propagation in the spacetime \bhsol, \bhuv.
A nice property of this spacetime is that the
string coupling
\eqn\gst{g_s=e^\Phi={1\over\cosh t}}
goes to zero at early $(t\to-\infty)$ and late
$(t\to+\infty$) times. This allows one to set up the
incoming and outgoing Hilbert spaces in regions where
interactions can be neglected; in these regions the
background is described by the timelike linear dilaton
solution discussed in subsection {\it 2.1}. In order to study
interactions, one has to extend the solution to all $t$.

In subsection {\it 3.1} we describe the $|t|\to\infty$ part of the
background. In subsection {\it 3.2} we describe the full solution,
which turns out to be a coset CFT. Some properties of
excitations in the spacetime \bhsol\ are discussed in
sections 4, 5.

\subsec{Timelike linear dilaton}

In this subsection we review some properties of the timelike
linear dilaton solution, an exact CFT describing string
propagation in flat spacetime with a varying string coupling,
given by \ggss. This solution is discussed \eg\ in
\refs{\MyersFV,\AntoniadisAA,\AntoniadisVI,\PolchinskiRQ}.

We start with the bosonic string. The worldsheet fields
$x^\mu=(t,x^i)$ are free, but the time dependence of the
dilaton is reflected in a modified worldsheet stress tensor,
\eqn\modstress{T(z)=-\partial x_i
\partial x^i+\partial t
\partial t-Q\partial^2 t~.}
The central charge is modified accordingly,
\eqn\central{
c=d+1-6Q^2~.
}
Since the total central charge must be equal to $26$ in the
bosonic string, one finds that $d$ and $Q$ are related,
\eqn\qdcor{6Q^2=d-25~.}
We see that asymptotically timelike linear dilaton solutions
are typically {\it supercritical}. One can replace
the $d$ free fields $x^i$ by an arbitrary unitary CFT, and
get more general timelike linear dilaton solutions in which
space is curved. For concreteness, we will mostly
discuss here the case of flat space.

Since the worldsheet theory is free and the string coupling is
small, it is not difficult to construct the physical states.
As in flat spacetime with constant dilaton, states are labeled
by momentum $p_\mu$ and an oscillator contribution. The corresponding
vertex operators have the form
\eqn\vertope{V=e^{ip_\mu x^\mu}
P_N(\partial x^\mu,\bar\partial x^\mu,\cdots)~,}
where $P_N$ is a polynomial in derivatives of the worldsheet
fields $x^\mu$ of total (left and right) scaling dimension $N$,
and we have taken the zero mode part of the vertex
operator to be a plane
wave. Physical states correspond to Virasoro primaries
of the form \vertope\ with scaling dimension one.
{}From \modstress, it follows that the scaling dimension of $V$ \vertope\ is
\eqn\lzero{
L_0={1\over 4}p_i^2-{1\over 4}p_0(p_0-2iQ)+N~.
}
The non-standard
contribution of $p_0$ to the scaling dimension is easy to
understand. In string theory, the (zero mode parts of)
vertex operators for emission of string modes have in
general the form
\eqn\verwave{V=g_s\Psi~,}
where $\Psi$ is the wavefunction of the state, and $g_s$
the string coupling. Usually, the factor of $g_s$ in \verwave\
can be neglected since it is constant, but here it is
time-dependent \ggss\ and therefore needs to be retained.
The vertex operator \vertope\ corresponds to the wavefunction
\eqn\psiwave{\Psi(\vec x, t)=e^{i\vec p\cdot \vec x}e^{i(p_0-iQ)t}~.}
Thus, the energy is
\eqn\zeromom{
E=p_0-iQ
}
and \lzero\ reads
\eqn\lzeromodif{L_0={1\over 4}\left(\vec p^2-E^2-Q^2\right)+N~.
}
The mass shell condition is
\eqn\masss{m_{\rm eff}^2=E^2-\vec p^2=4(N-1)-Q^2~.}
The effect of the linear dilaton is a downward shift
of all masses squared by $Q^2$. Denoting by $m^2$ the
combination
\eqn\mflat{m^2=4(N-1)~,}
as in the constant dilaton case, we have
\eqn\meffec{m_{\rm eff}^2=m^2-Q^2~.}
For example, the graviton behaves as a particle of
mass $m_{\rm eff}^2=-Q^2$ in the linear dilaton background.

The negative mass shift \masss\ leads to the fact that
there is a finite range of positive masses squared,
$0<m^2<Q^2$, which are effectively tachyonic. 
We will return to this issue in section~5.
For now, we compute the torus partition sum,
\eqn\partfion{
Z(\tau)={\rm Tr}(q^{L_0-{c\over 24}}\bar q^{\tilde L_0-{c\over 24}})~,
}
which summarizes the multiplicities of states at different
mass levels.
Here $q=e^{2\pi i\tau}$ and $\bar q=e^{-2\pi i\bar\tau}$.
On a Euclidean worldsheet, $\tau$ is complex, and
$\bar\tau=\tau^*$. On a Minkowski torus, $\tau$
and $\bar\tau$ are independent real numbers.

It is convenient to perform the calculation in the
covariant formalism. One finds
\eqn\partfionbos{
Z_{d+1}^{\rm bos}(\tau)=V_{d+1}\int {d^{d+1}p\over (2\pi)^{d+1}}
(q\bar q)^{-{d+1-6Q^2\over 24}+{1\over 4}
(\vec p^2-p_0^2+2iQp_0)}
(\sum_{\rm osc}q^N\bar q^{\tilde N})\ Z_{\rm ghost}~.
}
Here, $V_{d+1}$ is the volume of $d+1$ dimensional spacetime.
Completing the square in the exponent and rewriting the
resulting amplitude in terms of $E$ \zeromom, we find
\eqn\partfionbosbis{
Z_{d+1}^{\rm bos}(\tau)=V_{d+1}
\int {d^dp\over(2\pi)^d}
\int {dE\over2\pi} (q\bar q)^{-{d+1\over 24}+{1\over4}
(\vec p^2-E^2)}(\sum_{\rm osc}q^N\bar q^{\tilde N})\ Z_{\rm ghost}
~.
}
The string amplitude involves also an integral over $\tau$, $\bar\tau$.
Since spacetime has Lorentzian signature, one must take the worldsheet
to be Lorentzian as well.\foot{For recent discussions of Lorentzian
and Euclidean worldsheet and spacetime, see \refs{\BalasubramanianRY,
\LiuFT}.} The $\tau$ integral is then oscillatory, and
one needs to supply an $i\epsilon$ prescription to compute it. It is
possible to deform the contours of integration over $\tau$ and $E$
in \partfionbos\ such that $\tau$ becomes the modulus of a Euclidean torus,
and $E$ is integrated over the imaginary axis. This is a standard procedure
which is described \eg\ in \PolchinskiRQ, page 83.
The final answer one obtains is
\eqn\finbosz{
Z_{d+1}^{\rm bos}=\int{d^2\tau\over4\tau_2}
Z_{d+1}^{\rm bos}(\tau)=iV_{d+1}\int{d^2\tau\over16\pi^2\tau_2^2}
(4\pi^2\tau_2)^{-{(d-1)\over2}}
(\eta\bar\eta)^{1-d}~.}

Note that
the partition sum \finbosz\ is proportional to the volume of time. This
is natural from the spacetime point of view, since $Z_{d+1}$ is a trace
over free string states; it receives contributions from the infinite time
period during which the string coupling is very small, and the physics
is effectively time translation invariant. From the worldsheet
perspective, the factor of the volume of time arises since the linear
dilaton term in the worldsheet
Lagrangian, which breaks time translation invariance, vanishes on
the torus (it is proportional to $\hat R$, the curvature of the worldsheet,
which vanishes in this case). Thus, this amplitude is invariant under
time translations.

The negative mass shift \masss\ is manifest in the partition sum \finbosz.
{}From the form of the Dedekind $\eta$ function one finds that level matched
states with oscillator number $N$ contribute terms that go like
\eqn\levmatch{(q\bar q)^{-{d-1\over 24}+N}= (q\bar q)^{{1\over4}m_{\rm
eff}^2}~,}
where we used \qdcor, \masss. In particular, states with negative
$m_{\rm eff}^2$
contribute negative powers of $q\bar q$, a fact that will be
important in section~5, when we discuss infrared divergences.

\lref\KutasovSV{
D.~Kutasov and N.~Seiberg,
``Number Of Degrees Of Freedom, Density Of States
And Tachyons In String Theory And CFT,''
Nucl.\ Phys.\ B {\bf 358}, 600 (1991).
}

\lref\PolchinskiRR{
J.~Polchinski,
``String Theory. Vol. 2: Superstring Theory And Beyond,''
{\it  Cambridge, UK: Univ. Pr. (1998) 531 p}.
}

It is also worth pointing out that the negative mass
shift $-Q^2$ in \masss\ is related to another
interesting physical property of the model, its high
energy density of states. It was shown in \KutasovSV\
(see also \KutasovPV)
that in string theory there is a close relation between
infrared instabilities and the high energy density of
states (a UV-IR relation). The timelike linear dilaton
background discussed here is an example of this. Since the
model is supercritical, \ie\ the number of space dimensions is
larger than twenty five (see \qdcor), the high energy density
of states implicit in \finbosz\ is larger than that of
critical bosonic string theory. Correspondingly, the
infrared instability of the model is more severe than
that of $25+1$ dimensional bosonic string theory due to
the downward shifts of the masses of low lying states \masss.

We next move on to type II string theory in the timelike linear dilaton
background. Much of the discussion above goes through. The worldsheet
theory is now superconformal. The superconformal generators are
\eqn\modsuper{\eqalign{
T_B&=-\partial x_i\partial x^i+\partial t\partial t
-Q\partial^2 t-\half\psi^\mu\partial\psi_\mu~;\cr
T_F&=i\sqrt{2}(\psi_i\partial x^i-
\psi_t\partial t+Q\partial\psi_t)~.\cr
}}
The central charge is
\eqn\chat{\hat c={2\over3}c=d+1-4Q^2~.}
Setting $\hat c=10$ for criticality, we find
\eqn\dcond{4Q^2=d-9~.}
The main new issue in the type II case concerns the chiral
GSO projection. Modular invariance and spin-statistics
give rise to strong constraints that need not have
solutions generically. We will not attempt to classify
all possible solutions, but will illustrate the resulting
structures with an example from \AntoniadisVI.

A natural choice of a GSO projection is a separate
$(-)^{F_L}$ and $(-)^{F_R}$ projection on the left
and right movers, as in the critical superstring.
This gives rise to the partition sum\foot{We use
the notations of \PolchinskiRR.}
\eqn\iipart{Z^{II}_{d+1}=
iV_{d+1}\int {d^2\tau\over16\pi^2\tau_2^2}Z_x^{d-1}Z_\psi^+
\bar Z_\psi^\pm~,}
where
\eqn\zzxx{Z_x=(4\pi^2\tau_2)^{-\half}|\eta(\tau)|^{-2}}
and
\eqn\zpsipm{Z_\psi^\pm=\half\left[Z_0^0(\tau)^{d-1\over2}-
Z_1^0(\tau)^{d-1\over2}-Z_0^1(\tau)^{d-1\over2}\mp
Z_1^1(\tau)^{d-1\over2}\right]~.}
The choice of $\pm$ distinguishes between type IIA and
type IIB string theory. A peculiar feature of \iipart\
-- \zpsipm\ is that the partition sum is only modular
invariant in $d+1=10+16m$ dimensions, $m=1,2,\ldots$.
The resulting theory is not spacetime supersymmetric,
which can be seen by noting that the worldsheet fermion
contribution \zpsipm\ is not zero for $d+1\not=10$.
For example, in the twenty six dimensional non-critical
superstring vacuum,
\eqn\idtwentysix{Z_\psi^\pm=
\half\left[Z_0^0(\tau)^{12}-
Z_1^0(\tau)^{12}-Z_0^1(\tau)^{12}\mp
Z_1^1(\tau)^{12}\right]=24~.}
Thus, there are large cancellations between (spacetime)
bosons and fermions in the spectrum, but they are
incomplete. The full partition sum in this case is
\eqn\Ztwo{
Z^{II}_{d+1=26}(\tau)=(24)^2 Z^{24}_x(\tau)~.
}

\noindent
So far we discussed some properties of the spectrum of
string theory in the timelike linear dilaton background.
As explained in the beginning of this section, this is
sufficient for studying the spectrum of asymptotic states
in the full classical string theory in the background \bhuv,
because the background approaches a timelike linear dilaton
one at $|t|\to\infty$.

To study interactions in general, one needs to understand
the full background \bhuv\ in string theory. Nevertheless,
some interactions can be studied using the linear dilaton
CFT. As is clear from figure 2, one expects two kinds of
interactions in the background \bhuv. One kind
involves processes that take place in the part of the geometry
far from the big bang/big crunch region, $t\simeq 0$.
Such processes, which are known as ``bulk amplitudes''
(see \eg\ \refs{\DiFrancescoUD,\KutasovPV} for a discussion)
can be studied by computing Shapiro-Virasoro type amplitudes
in the timelike linear dilaton background. They are proportional
to the volume of the (infinite) time interval well before or after
the big bang or big crunch. We will not discuss them here,
since they are rather standard, and do not shed light on the physics
associated with the singularity.

Of more interest is the second kind of interactions, which occur
near $t=0$, and thus are given by amplitudes that do not go like
the volume of time. To study these, one needs to understand the
full background \bhuv, to which we turn next.

\subsec{SL(2,\IR)/U(1)}

\lref\BarsTI{
I.~Bars and K.~Sfetsos,
``Global analysis of new gravitational singularities in string and particle theories,''
Phys.\ Rev.\ D {\bf 46}, 4495 (1992)
[hep-th/9205037].
}
\lref\BarsSR{
I.~Bars and K.~Sfetsos,
``Conformally exact metric and dilaton in string theory on curved space-time,''
Phys.\ Rev.\ D {\bf 46}, 4510 (1992)
[hep-th/9206006].
}
We would like to understand the solution \bhuv\ in string
theory. The answer to this question is in fact known. As
shown in  \WittenYR, the coset CFT $SL(2,\IR)/U(1)$ where
one gauges a non-compact spacelike $U(1)$ in $SL(2)$, gives
rise to a sigma model in the background\foot{For early work
on generalizations to other cosets and applications to cosmology,
see \eg\ \refs{\TseytlinXK,\KounnasWC,\BarsTI,\BarsSR}.}
\eqn\cosetbg{
\eqalign{
ds^2&=-k {du\,dv\over1-uv}~;\cr
\Phi&=
-{1\over4}{\rm log}(1-uv)^2~.
}
}
$k$ is the level of the $SL(2)$ affine Lie
algebra that enters the coset
construction.\foot{Actually, \cosetbg\ with
$-\infty<u,v<\infty$ describes $PSL(2,\IR)/U(1)$.
$SL(2,\IR)/U(1)$ is a double cover of this space.
We will be mostly interested in the single cover,
\ie\ the $PSL(2)$ case.}
For bosonic strings, the background \cosetbg\ is only valid for
large $k$. The $1/k$ (or $\alpha'$) corrections are known
\refs{\DijkgraafBA,\JackMK}. For the worldsheet supersymmetric
version, which is the case of interest to us here,
the background \cosetbg\ turns out to be exact
\refs{\TseytlinHT,\JackMK}, with
$k$ being the total level of $\widehat{SL(2)}$ (which
receives a contribution of $k+2$ from a bosonic $SL(2)$ WZW
model and a contribution of $-2$ from three free fermions
in the adjoint of $SL(2)$). The central charge of the resulting
superconformal field theory is
\eqn\chatsl{\hat c=2+{4\over k}~.}
Comparing \bhuv\ to \cosetbg, we see that the generalized Milne
universe described in subsection {\it 2.3} corresponds to a negative
level $SL(2)/U(1)$,
\eqn\kQmatch{{1\over k}=-Q^2~.}
Thus, the $SL(2)/U(1)$ CFT has central charge $\hat c<2$.
It is also known that in this case the $N=1$ superconformal
symmetry of the coset model is enhanced to $N=2$. For
unitary CFT's these two facts would mean that the model
must be an $N=2$ minimal model, and $k$ must be integer.
However, in our case there is no reason to impose unitarity
on the CFT, since the $SL(2)/U(1)$ directions include time.

The background \bhuv\ can be thought of as a double Wick
rotated version of the two dimensional black hole. This
is reflected in the Penrose diagram of figure 1. The
region outside the horizon of the black hole corresponds
in our case to the early time epoch (region $I$ in
figure 1), while the other asymptotic region in the black
hole case corresponds to the late time region $II$.
The regions between the horizon and singularity of the black
and white holes correspond in our case to regions
$III$ and $IV$. The singularities at $uv=1$ appear
timelike in regions $III$, $IV$. Finally, the regions
behind the black and white hole singularities, $V$ and
$VI$, give rise to additional asymptotic regions.

In section 2 we discussed a background obtained from \bhuv\ by the
identification \uvid. This spacetime can be obtained from
$SL(2,\IR)/U(1)$ by further modding out by a discrete group.
In terms of the $SL(2,\IR)$ matrices $g$, the background
\bhuv\ is obtained by gauging the continuous symmetry
\eqn\ggbb{g\to e^{\rho\sigma_3}ge^{\rho\sigma_3}~.}
The identified generalized Milne universe discussed in
subsection {\it 2.3} is obtained by further gauging the
discrete subgroup isomorphic to $\IZ$, which is generated
by
\eqn\ggaa{g\to e^{\tau\sigma_3}ge^{-\tau\sigma_3}~,}
where $\tau=\pi R$ (note the factor of two relative to \uvid).

A note on T-duality: when constructing the $SL(2,\IR)/U(1)$
model, one has a choice of gauging either the vector symmetry
\ggbb, or the axial symmetry $g\to \exp(\rho\sigma_3)g
\exp(-\rho\sigma_3)$. The CFT is self-dual under this
replacement \refs{\GiveonSY,\DijkgraafBA},
but duality exchanges regions $I$ and $VI$,
as well as regions $II$ and $V$. Regions $III$ and $IV$
are invariant. This will play a role in our discussion in section~4.

In the next section we turn to the study of excitations
propagating in the background \bhuv. We will use the
description of the singular spacetime of figure 2 as a
coset of $PSL(2)$, to continue wavefunctions between
different regions separated by cosmological singularities,
and study the resulting S-matrix.

\newsec{Scalar perturbations}

We would like to analyze the propagation of small perturbations
in the background \bhsol, \bhuv. For concreteness, we will consider
a minimally coupled scalar field $T$ of mass $m$, which could (for
example) be one of the perturbative string modes described in the
previous section. We will focus on the non-trivial
two dimensional part of the spacetime. The other $d-1$ dimensions
are flat; it is easy to incorporate them into the discussion.

The contribution of $T$ to the action \action\ is
\eqn\stach{S_T=
{1\over 2\kappa^2} \int dxdt\sqrt{-g}
e^{-2\Phi}\left(-g^{\mu\nu} \partial_\mu
T \partial_\nu T- m^2T^2\right)~,
}
where we are neglecting self-interaction terms, as well
as interactions with other fields, and focus on free
propagation in the geometry \bhsol, \bhuv.

Since the background \bhsol\ is translation invariant
in $x$, we will consider modes with a given value of
spatial momentum,
\eqn\planetach{T(t,x) = T(t) e^{ipx}+{\rm c.c.}}
Substituting this ansatz into the action \stach, and
adding \mini, one finds
\eqn\minitach{S_g+S_T=
-{1\over 2\kappa^2}
\int dt \sqrt{-g_{00}}{e^{-\phi}\left[
g^{00}(\dot\alpha^2 -\dot\phi^2+\dot T^2)
+2\Lambda+m^2T^2+
e^{-2\alpha(t)}p^2 T^2\right]}~.
}
The resulting equations of motion are (compare to \kassoln)
\eqn\tacheom{\eqalign{
&{\partial\over\partial t}\left(e^{-\phi}\dot\alpha\right)=
p^2e^{-2\alpha(t)} T^2~;\cr
&\ddot\phi = \dot\phi^2 - 2 \Lambda-
(m^2+p^2e^{-2\alpha(t)})T^2 = \dot\alpha^2+\dot T^2~;\cr
&\ddot{T} - \dot{\phi}\dot{T} + (m^2 + p^2 e^{-2\alpha(t)})T = 0~.
\cr
}}
If the scalar field $T$ is small everywhere, one can neglect
the $O(T^2)$ back-reaction of the metric and dilaton, which is
summarized by the first two equations in \tacheom, and solve
the linearized equation of motion of $T$ given by the last line
of \tacheom, in the fixed background \bhmet, \bhuv.

Substituting the values of $\alpha$ and $\phi$ corresponding
to the solution~{\bhmet, \bhdil}, and defining
\eqn\newdefs{z=-{\rm sinh}^2(Qt),\;\; M={m\over Q}~,}
brings the last line of \tacheom\ to the form
\eqn\hgf{
z(1-z) T'' + (1 - 2z) T' -{1\over4} \bigl(M^2 -
p^2 {1-z\over z}\bigr) T
= 0~,}
where primes denote derivatives with respect to $z$. The
asymptotic past and future regions, with $|t|\to\infty$,
correspond to
\eqn\zlimit{z\simeq -(1/4)\exp(2Q|t|)\to-\infty~.}
In this limit \hgf\ simplifies:
\eqn\asymhgf{-z^2T''-2zT'-{1\over4}(M^2+p^2)T=0~,}
with the solution
\eqn\solnasymp{\eqalign{
T=&(-z)^\lambda~;\cr
\lambda=&\half\left(-1\pm \sqrt{1-M^2-p^2}\right)~.\cr
}}
For $M^2+p^2>1$, this corresponds to
plane waves with energy
\eqn\EQQ{E=Q\sqrt{M^2+p^2-1}~.}
For smaller masses one gets growing and decaying wavefunctions,
as discussed in subsection~{\it 3.1}. The bound on $M$ in
\solnasymp\ is the same as the one that follows from \meffec.
In this section we will concentrate on the plane waves (with
$E\in \IR$). We will return to the growing solutions in section~5.

Equation \hgf\ can be solved in terms of hypergeometric
functions, but it is more instructive to proceed by using the
construction of the space \bhuv\ as a coset of $PSL(2,\IR)$. This
provides a unified description of the behavior of the wavefunctions
in all six regions in figures 1,2.

One starts by constructing eigenfunctions of the Laplacian
on $PSL(2,\IR)$, in which the $U(1)_L\times U(1)_R$ symmetries
$g\to \exp(\rho_L\sigma_3)g\exp(\rho_R\sigma_3)$
are diagonal. This is described in \Vilenkin\
(see \ElitzurRT\ for a review and an application in a related
context), so we will be brief.
The eigenfunctions of the Laplacian are matrix
elements of group elements between states belonging
to representations of $PSL(2,\IR)$. The scattering states
correspond to matrix elements in the principal continuous
series. States in these representations are labeled by
three quantum numbers, $j$; $m$; $\pm$. The quantum number
$j=-\half+is$, $s\in \IR$, is related to the value of
the Casimir, $j(j+1)$. The label $m$ determines the
eigenvalue under the non-compact $U(1)$ symmetry
$\exp(\alpha\sigma_3)$; for unitary representations,
this eigenvalue is $\exp(2im\alpha)$, with $m\in \IR$. The
third quantum number is a discrete label which takes two values.

A generic group element $g\in PSL(2,\IR)$ (with all
elements of the $2\times 2$ matrix $g$ non-vanishing)
can be written as
\eqn\grepr{g(\alpha,\beta,\theta;\epsilon_2,\delta)=
e^{\alpha\sigma_3}(i\sigma_2)^{\epsilon_2}
g_\delta(\theta)e^{\beta\sigma_3}~,}
where $\epsilon_2=0,1$; $\delta=I,II,IV$, and
\eqn\gI{g_{IV}=
\left(\matrix{
\cos\theta&-\sin\theta\cr
\sin\theta&\cos\theta\cr}\right);\qquad 0\le\theta\le{\pi\over2}~,}
\eqn\gone{g_I=g_{II}^{-1}=
\left(\matrix{
\cosh\theta&\sinh\theta\cr
\sinh\theta&\cosh\theta\cr}\right);\qquad 0\le\theta<\infty~.}
The non-vanishing matrix elements of $g$ \grepr\ in the
principal continuous series are
\eqn\matelem{
K_{\pm\pm}(\lambda,\mu;j;g)\equiv
\langle j, m,\pm|g|j,\bar m,\pm\rangle=
e^{2i(m\alpha+\bar m\beta)}
\langle j, m,\pm|(i\sigma_2)^{\epsilon_2}
g_\delta(\theta)|j,\bar m,\pm\rangle~,
}
where
\eqn\lmmu{
\lambda\equiv -im-j;\;\; \mu\equiv -i\bar m-j;\;\;j=-\half+is~.}
The labels $m$, $\bar m$, $s$ take arbitrary real values.
The functions \matelem\ appear in \Vilenkin\ (see also
\ElitzurRT). We will present some of them later.

Given the eigenfunctions of the Laplacian on $PSL(2,\IR)$,
we can find the wavefunctions on the coset obtained by gauging
\ggbb, by restricting to gauge invariant wavefunctions.
Since the parameters $\alpha$, $\beta$ in \grepr\ transform under
\ggbb\ as $\alpha\to\alpha+\rho$, $\beta\to\beta+\rho$,
the invariant wavefunctions are those with $m=-\bar m$. Further
modding out by the identification \ggaa\ leads to quantization of $m$,
$mR\in \half\IZ$. Also, twisted sectors appear, in which $m\not=-\bar m$.
These correspond to winding modes on the coset manifold.

Thus, the matrix elements \matelem\ give rise to
solutions of the wave equation on the spacetime \bhuv.
Different regions in figures 1,2 correspond\foot{This is described
in more detail in \ElitzurRT. The regions in figure 1 are
related to those in figure~2 of \ElitzurRT\ as follows. Our regions
$(I,II,III,IV,V,VI)$ correspond to regions $(1,1',II,I,2',4)$
in \ElitzurRT.} to different types of $PSL(2,\IR)$
matrices $g$, \grepr. For example, region I in figure 1
corresponds to $\epsilon_2=0$, $\delta=I$ in \grepr.
The coordinates on region I in \grepr, $\alpha-\beta$ and
$\theta$, are related to the coordinates $x$ and $t$ in
\bhsol\ as follows:
\eqn\coordmap{\eqalign{
x&\leftrightarrow\alpha-\beta~; \cr
t&\leftrightarrow\theta ~. \cr
}}
The two independent solutions of \hgf\ in region $I$
can be taken to be
\Vilenkin
\eqn\kplusone{
T^{(+)}_I(t)=K_{++}(\lambda,\mu;j;g_I)=
{1\over2\pi i}B(\lambda, -\lambda - 2j)
{(1-z)^{j + {\lambda + \mu\over 2}}\over (-z)^{\lambda + \mu\over 2}}
F(\lambda,\mu;-2j;{1\over z})}
and
\eqn\kminusone{\eqalign{
T^{(-)}_I(t)&=K_{--}(\lambda,\mu;j;g_I)\cr
&= {1\over2\pi i}B(1-\mu,\mu+2j+1)
{(1-z)^{j+{\lambda+\mu\over 2}}\over (-z)^{2j+1+{\lambda+\mu\over 2}}}
F(\lambda+2j+1,\mu+2j+1;2j+2;{1\over z})~,\cr
}}
where $z$ is given in \newdefs,
$B(a,b)$ is the Euler Beta function
\eqn\bfff{B(a,b)={\Gamma(a)\Gamma(b)\over\Gamma(a+b)}~,}
and $F(a,b;c;x)$ is the hypergeometric function ${}_2F_1$.
By comparing the asymptotic behavior of the two solutions
\kplusone, \kminusone\ to \solnasymp, one can match the
parameters $s, m$ in \lmmu\ with those in \solnasymp:
\eqn\parmatch{\eqalign{
m=&-\bar m={p\over2}~;\cr
2s=&\sqrt{M^2+p^2-1}=E/Q~.\cr
}}
For the case of compact $x$, there are both momentum and
winding modes, with the spectrum
\eqn\momwin{
\eqalign{
m=&\half({n\over R}+kwR)~;\cr
\bar m=&\half(-{n\over R}+kwR)~.\cr
}}
Here, $n,w\in\IZ$ are the momentum and winding,
respectively, while $k$ is the level of $\widehat{SL(2)}$.

Since the wavefunctions $K_{\pm\pm}$ are given by
matrix elements on $PSL(2,\IR)$, \matelem, they
are uniquely defined in all regions in figure 1.
For example, the solution $T^{(+)}_I(t)$, which in region $I$ is given by
\kplusone, can be determined in regions $II$, $VI$,
by using the fact that \refs{\Vilenkin,\ElitzurRT}
\eqn\konetwo{\eqalign{
K_{++}(\lambda,\mu;j;g_{II})=& K_{--}(\lambda,\mu;j;g_{I})~;\cr
K_{++}(\lambda,\mu;j;g_{VI})=& 0~,\cr
}}
where $g_{II}=g_I^{-1}$, as in \gone, and $g_{VI}=-g_Ii\sigma_2$.
Another region that will be of interest below is region $V$, where
\eqn\kppfive{\eqalign{
K_{++}(\lambda,\mu;j;g_V)&= {1\over 2 \pi i}(1-z)^{\mu - \lambda\over 2}\cr
\times&\bigl[B(-2j - \lambda, 1- \mu)(-z)^{-{\lambda+\mu+2j\over 2}}
F(-\lambda -  2j,1-\lambda;- \mu-\lambda - 2j+ 1;z)\cr
&+B(\lambda, \mu+2j+1)(-z)^{\lambda+\mu+2j\over 2}
F(\mu,\mu+2j+ 1;\mu+\lambda + 2j+ 1;z)\bigr]\cr
}}
with  $g_V=i\sigma_2 g_I$.

Similarly, one can compute the other three eigenfunctions of
the Laplacian $K_{+-}$, $K_{-+}$, $K_{--}$ in all six regions
of figure 1. Of course, in any given region, only two of the
four functions $K_{\pm\pm}$ are independent. The others can
be expressed as linear combinations of any linearly independent
two (and in some cases vanish, such as $K_{++}$ in region $VI$ \konetwo).
We will give some additional examples below.

We are now ready to return to the problem of quantum field theory in
the background \bhuv. Naively, in a spacetime with metric \bhmet,
one would expect to specify an initial state at $t\to-\infty$
and calculate the amplitude for it to propagate to some particular
final state at $t\to+\infty$. However, extending the spacetime
as in \bhuv\ and figures 1, 2, it is clear that the initial time
surface actually contains two independent components, corresponding
to early times in regions $I$ (which is covered by the original
coordinates \bhmet) and $VI$. Similarly, the out-region is the
union of late times in regions $II$ and $V$. Thus, the incoming
Hilbert space is expected to be the direct product of the Hilbert
spaces corresponding to regions $I$ and $VI$, while the outgoing
Hilbert space is expected to be the direct product of Hilbert spaces
corresponding to regions $II$ and $V$. This picture is further supported
by the fact that regions $I$ and $VI$ are exchanged by T-duality
\refs{\GiveonSY,\DijkgraafBA}, and so are regions $II$ and $V$.

To quantize the field $T$ \stach, we would like to expand it in a complete
set of orthonormal mode solutions. In the early and late time regimes
$|t|\to\infty$, the metric becomes Minkowski \timedil, and the string
coupling goes to zero and so can be neglected. Therefore, it is natural
to treat $i\partial/\partial t$ as the timelike Killing vector with respect
to which one measures energies (as we have done in section 3). There are
two natural sets of modes in terms of which the quantum field $T$ can
be expanded, corresponding to the early and late time regimes, respectively.
We next construct these modes.

At early times, we are looking for two types
of solutions. The first type corresponds to wavefunctions which vanish
in region $VI$, and contain only positive frequency modes as $t\to-\infty$
in region $I$. Wavefunctions of the second type vanish in region $I$ and
contain only positive frequency modes at early times in region $VI$.

The mode solutions of the first type
are $K_{++}(\lambda,\mu;j;g)$ with $s>0$ in \lmmu. This solution
indeed vanishes in region $VI$ (see \konetwo), and its
$t\to-\infty$ behavior in region $I$ is determined by
\zlimit, \kplusone:
\eqn\tplusas{
K_{++}(\lambda,\mu;j;g_I)(t\to-\infty)\simeq {1\over2\pi i}
B(\lambda,-\lambda-2j)(-z)^j\simeq
{1\over2\pi i}B(\lambda,-\lambda-2j)4^{-j}e^{-2Qtj}~.}
Substituting $j=-\half+is$ we have
\eqn\plssss{K_{++}(\lambda,\mu;j;g_I)(t\to-\infty)\simeq {1\over2\pi i}
B(\lambda,-\lambda-2j)
4^{-j}e^{Qt}e^{-2iQst}~.}
The factor $\exp(Qt)$ is the string coupling
(see \timedil, \verwave), while
$\exp(-2iQst)=\exp(-iE t)$ (see \parmatch).
Thus, $K_{++}(\lambda,\mu;j;g)$ contains only
positive frequency modes at early times in region $I$.

Similarly, the positive energy solutions of the second type
correspond to  $K_{+-}(\lambda,\mu, j, g)$.
One can show \Vilenkin\ that $K_{+-}$ satisfies the following
properties:
\eqn\proppm{\eqalign{
K_{+-}(\lambda,\mu,j,g_I)&=K_{++}(-\lambda-2j,\mu,j,g_{VI})=0~;\cr
K_{+-}(\lambda,\mu,j,g_{VI})&=K_{++}(-\lambda-2j,\mu,j,g_{I})~,\cr
}}
where we recall that $g_{VI}=-g_Ii\sigma_2$.
The first line of \proppm\ shows that the
mode $K_{+-}(t\to-\infty)$ lives purely in region
$VI$. The second line gives the profile of $K_{+-}$
in region $VI$, and in particular establishes
that it corresponds to a positive energy solution
as $t\to-\infty$ in this region.

One can expand the quantum field $T$ in the above modes,            
\eqn\texpand{\eqalign{
N\;T(\lambda,\mu)= &a_I(\lambda,\mu)K_{++}(\lambda,\mu;j;g)
+a^\dagger_I(\lambda,\mu)K^*_{++}(\lambda,\mu;j;g)\cr
+&a_{VI}(\lambda,\mu)K_{+-}(\lambda,\mu;j;g)+
a^\dagger_{VI}(\lambda,\mu)K^*_{+-}(\lambda,\mu;j;g)~,\cr}}
where $N$ is a normalization factor, which can be read off
from \plssss. The vacuum $|0\rangle_{\rm in}$ is defined
to be the state annihilated by the annihilation
operators in regions $I$ and $VI$, $a_I$ and $a_{VI}$.
The operators $a^\dagger_I$ and $a^\dagger_{VI}$
create particles in regions $I$ and $VI$,
respectively. The fact that the modes  $K_{++}$ and
$K_{+-}$,  corresponding to the two in-regions $I$ and $VI$,
are orthogonal in the Klein-Gordon norm (because
either one or the other vanishes at all points on an early time
Cauchy surface),  implies that the creation and annihilation operators
$a_I, a^\dagger_I$ commute with $a_{VI}, a^\dagger_{VI}$.

At late times, it is more natural to choose modes that vanish
in one of the out-regions and are purely positive frequency in
the other one. Using results in \Vilenkin\ one finds that
the solution of the wave equation that vanishes in region $V$
and has purely positive energy in region $II$ is
$K_{--}^*(-\lambda-2j,-\mu-2j;j;g)$, while the positive energy solution
in region $V$ which vanishes in region $II$ is
$K^*_{-+}(-\lambda-2j,-\mu-2j;j;g)$. Therefore, the expansion appropriate
for late times is
\eqn\texpandtwo{\eqalign{
N\;T(\lambda,\mu)= &a_{II}(\lambda,\mu)K^*_{--}(-\lambda-2j,-\mu-2j;j;g)
+a^\dagger_{II}(\lambda,\mu)K_{--}(-\lambda-2j,-\mu-2j;j;g)\cr
+&a_{V}(\lambda,\mu)K^*_{-+}(-\lambda-2j,-\mu-2j;j;g)+
a^\dagger_{V}(\lambda,\mu)K_{-+}(-\lambda-2j,-\mu-2j;j;g)~.\cr}}
The out-vacuum $|0\rangle_{\rm out}$ is by definition annihilated
by the annihilation operators $a_{II}$ and $a_{V}$.

To determine the relation between the incoming
and outgoing vacua, one needs to expand the outgoing
modes $K_{--}$, $K_{-+}$ \texpandtwo\ in terms of the
incoming ones $K_{++}$, $K_{+-}$ \texpand. This can be
done, for example, by matching the asymptotic behaviors
at early times (in regions $I$ and $VI$).
In region $I$, the asymptotic behavior of the mode
solutions is
\eqn\asympt{\eqalign{
K_{++}(\lambda,\mu;j;g_I) & \sim {1\over 2\pi
i}B(\lambda,-\lambda-2j) (-z)^j~;\cr
K_{+-}(\lambda,\mu;j;g_I) &=0~;\cr
K_{--}(\lambda,\mu;j;g_I) & \sim {1\over 2\pi
i}B(1-\mu,\mu+2j+1) (-z)^{-j-1}~;\cr
K_{-+}(\lambda,\mu;j;g_I) & \sim {1\over 2\pi
i}[B(2j+1,-2j-\lambda)+B(2j+1,\lambda)] (-z)^{j}\cr
&+{1\over 2\pi
i}[B(-2j-1,\mu+2j+1)+B(-2j-1,1-\mu)] (-z)^{-j-1}~.\cr
}
}
In region~VI, it is
\eqn\asymptsix{\eqalign{
K_{+-}(\lambda,\mu;j;g_{VI}) & \sim {1\over 2\pi
i}B(\lambda,-\lambda-2j) (-z)^j~;\cr
K_{++}(\lambda,\mu;j;g_{VI}) &=0~;\cr
K_{-+}(\lambda,\mu;j;g_{VI}) & \sim {1\over 2\pi
i}B(1-\mu,\mu+2j+1) (-z)^{-j-1}~;\cr
K_{--}(\lambda,\mu;j;g_{VI}) & \sim {1\over 2\pi
i}[B(2j+1,-2j-\lambda)+B(2j+1,\lambda)] (-z)^{j}\cr
&+{1\over 2\pi
i}[B(-2j-1,\mu+2j+1)+B(-2j-1,1-\mu)] (-z)^{-j-1}~.\cr
}
}
{}From \asympt\ and \asymptsix, one obtains
\eqn\bogol{
\pmatrix{K_{--}\cr K_{-+}\cr K^*_{--}\cr K^*_{-+}}=
\pmatrix{A&C&0&B\cr
         C&A&B&0\cr
         0&B^*&A^*&C^*\cr
         B^*&0&C^*&A^*}
\pmatrix{K^*_{++}\cr K^*_{+-}\cr K_{++}\cr K_{+-}}~,
}
with
\eqn\abc{
\eqalign{
A&=-{B(1-\mu,\mu+2j+1)\over B(1-\lambda,\lambda+2j+1)}~;\cr
B&={B(2j+1,-2j-\lambda)+B(2j+1,\lambda)\over
B(\lambda,-\lambda-2j)}~;\cr
C&=-{B(-2j-1,\mu+2j+1)+B(-2j-1,1-\mu)\over
B(1-\lambda,\lambda+2j+1)}~.
}
}
{}From \texpand, \texpandtwo\ and \bogol, the following Bogolubov
transformation follows:
\eqn\bogola{
\pmatrix{a^\dagger_{I}\cr a^\dagger_{VI}\cr a_{I}\cr a_{VI}}
=\pmatrix{A&C&0&B^*\cr
         C&A&B^*&0\cr
         0&B&A^*&C^*\cr
         B&0&C^*&A^*}
\pmatrix{a^\dagger_{II}\cr a^\dagger_{V}\cr a_{II}\cr a_V}~.
}
The Bogolubov coefficients in \bogola\ satisfy the
relations
\eqn\relbogol{\eqalign{
|A|^2+|C|^2-|B|^2=&1~;\cr
AC^*+A^*C=&0~,\cr
}}
which are equivalent to equations (3.39) and (3.40) in
\BirrellIX. The relations \relbogol\ imply that \bogola\
can be inverted to
\eqn\invbogola{
\pmatrix{a^\dagger_{II}\cr a^\dagger_{V}\cr a_{II}\cr a_{V}}
=\pmatrix{A^*&C^*&0&-B^*\cr
         C^*&A^*&-B^*&0\cr
         0&-B&A&C\cr
         -B&0&C&A}
\pmatrix{a^\dagger_{I}\cr a^\dagger_{VI}\cr a_{I}\cr a_{VI}}~.
}
The Bogolubov coefficients can be used to determine some of
the S-matrix elements of the field $T$, as in \BirrellIX.
In particular, the fact that $B\neq0$ signals particle
creation (see \refs{\AharonyCX,\CornalbaNV} for  recent
discussions of particle creation in string theory).

The incoming vacuum evolves to an outgoing state which contains
\eqn\sssnnn{
{}_{\rm in}\langle 0|a^\dagger_{II}a_{II}|0\rangle_{\rm in}=
{}_{\rm in}\langle 0|a^\dagger_Va_V|0\rangle_{\rm in}=|B|^2}
particles in region $II$ as well as region $V$,
in the mode labeled by $\lambda, \mu, s$. Plugging
in the explicit form of $B$, given in \abc,
one finds that
\eqn\bsquared{|B|^2={\cosh^2\pi m\over\sinh^2\pi s}~,}
where we have used \lmmu; $m$ and $s$ are related to
the energy, momentum and winding via \parmatch\ and \momwin. 

We finish this section with a few comments. Our analysis
of Bogolubov coefficients leading to \bogola\ is strictly
speaking only valid in the large $|k|$ approximation.\foot{Recall
that $k$ is the level of $\widehat{SL(2)}$ entering the
construction (see subsection {\it 3.2}).}
While we have not studied the $1/k$ corrections in detail,
experience with $AdS_3$ \TeschnerFT\ suggests that the only
difference with respect to the classical analysis is the
appearance of a ``reflection coefficient'' relating
vertex operators corresponding to the representations
$|j\rangle$ and $|-j-1\rangle$. One expects in general to have
\eqn\ojref{\tilde\theta(-j-1)={\Gamma(1-{2j+1\over k})\over
\Gamma(1+{2j+1\over k})}\theta(j)~,}
where $\theta(j)$, $\tilde\theta(-j-1)$ are the quantum
vertex operators corresponding to the classical wavefunctions
which satisfy the $|k|\to\infty$ limit of \ojref,
$\tilde\theta(-j-1)=\theta(j)$.
This should be checked in more detail, but if true, it implies
that the $1/k$ corrections modify \bogola\ in the following
simple way: $A$ and $C$ \abc\ are multiplied by
$\Gamma(1-{2j+1\over k})/\Gamma(1+{2j+1\over k})$, while $B$
remains the same. Since for $j=-\half+is$, the correction factor
is a phase, it does not modify the consistency conditions \relbogol.

\lref\MaldacenaKM{
J.~M.~Maldacena and H.~Ooguri,
``Strings in AdS(3) and the SL(2,R) WZW model. III: Correlation  functions,''
hep-th/0111180.
}

\lref\GiveonTQ{
A.~Giveon and D.~Kutasov,
``Comments on double scaled little string theory,''
JHEP {\bf 0001}, 023 (2000)
[hep-th/9911039].
}

Another interesting extension of the analysis of this
section involves the calculation of higher point functions
in the cosmological spacetime \bhuv. This can in principle be
done by using results on $SL(2,\IR)$
\refs{\TeschnerFT,\MaldacenaKM}. For example, to calculate three
point functions one needs to perform a transform of the
$SL(2,\IR)$ correlation functions computed in \TeschnerFT.
A similar transform was discussed in the context of Little
String Theory in \GiveonTQ. It seems to be well behaved there;
more work is needed to establish whether the three point
functions are also well behaved here.

Four and higher point functions are of course of interest
as well, and one may hope that they can be understood
using the underlying $SL(2,\IR)$ affine Lie algebra structure.
Of particular interest are interactions that take place near
the singularities in figure 2. The amplitudes that are sensitive
to such interactions should receive contributions from principal
discrete series and degenerate representations of $SL(2)$, which
correspond to wavefunctions localized there, and can appear as
intermediate states in four and higher point functions.

\newsec{Growing  modes and infrared divergences in asymptotically timelike linear dilaton spacetimes}

In the previous section we studied perturbations
of the generalized Milne universe \bhuv\ which
correspond to particles with real energy $E$
\EQQ. As we noted after \EQQ, for sufficiently
low mass, $m^2<Q^2$, the energy is imaginary
(for low enough momentum), so that the wavefunctions
grow or decay exponentially with time.
In this section, we will briefly discuss such modes, 
first classically, and then at one loop.

\subsec{Classical analysis}

Consider the classical evolution of a scalar field $T$ with mass
$m$ in the linear dilaton background \timedil. For concreteness,
we will first discuss the early time epoch, during which $g_s=\exp(Qt)$
(for late times, one simply replaces $Q$ by $-Q$ in the following equations).
The Lagrangian \stach\ is proportional to
\eqn\lscalar{
\CL=\half e^{-2Qt} (-\eta^{\mu\nu}\partial_\mu T\partial_\nu T-m^2 T^2)~.
}
The equation of motion of $T$ is
\eqn\eom{
(-\eta^{\mu\nu}\partial_\mu\partial_\nu+2Q\partial_t+m^2) T=0~.
}
A basis of solutions is given by
\eqn\basissol{
T(t,\vec x)=e^{Qt}e^{\pm iEt+i\vec k\cdot\vec x}~,
}
with
\eqn\eeee{
E=\sqrt{\vec k^2+m^2-Q^2}~.
}
As discussed in section 3, it is natural to define the wavefunction
$\Psi$ to be the field $T$ with the factor of the coupling stripped
off. While this was described in the context of string theory, it
is very natural in field theory
as well \SeibergBJ. The resulting wavefunction,
\eqn\wwavefion{
\Psi(t,\vec x)=e^{-Qt}T(t,\vec x)= e^{\pm iEt+i\vec k\cdot\vec x}
}
has the property that for $\vec k^2+m^2-Q^2<0$, it is not a
plane wave, but an exponentially growing or decaying function
of $t$,
\eqn\expwave{\Psi(t,\vec x)=e^{\pm \omega t+i\vec k\cdot\vec x}}
with $\omega=-iE$ \eeee.

One consequence of this fact is the following.
Rewrite the Lagrangian 
\lscalar\ in terms of the wavefunction $\Psi$, by using 
\wwavefion. This is natural since at early times $\Psi$
is canonically normalized, while $T$ has a time dependent
kinetic term \lscalar. Omitting a total derivative which
does not influence the classical equations of motion,
one finds 
\eqn\lscpsi{
\CL=\half (-\eta^{\mu\nu}\partial_\mu \Psi\partial_\nu \Psi-
m_{\rm eff}^2 \Psi^2)~,
}
where $m_{\rm eff}$ is given by \meffec. This form makes it clear that
the the physics at very early times is invariant under time translations, 
and that fields with negative $m_{\rm eff}^2$ behave like tachyons
in flat space with constant dilaton. For example, the Hamiltonian
corresponding to \lscpsi,
\eqn\hamilt{
H=\half\int d^dx (\dot \Psi^2+(\nabla \Psi)^2+m_{\rm eff}^2\Psi^2)~,
}
is not constant for fields with $m_{\rm eff}^2+\vec k^2<0$: rather, it grows 
or decays exponentially with time. 

Since fields with $m_{\rm eff}^2<0$ look like tachyons \lscpsi,
it is natural to ask whether they lead to instabilities of the 
space. Classically, this is the questions whether as one evolves 
the growing wavefunctions \expwave\ forward in time, the self-interactions
of the field $T$ and its interactions with other fields, such as the
metric and dilaton, lead to large changes in the solution \bhuv\ even
if the initial perturbation is very small. 

Consider first a spacetime that approaches an asymptotically timelike
linear dilaton solution as $t\to+\infty$. The question whether such
a solution is stable against perturbations by modes with $m_{\rm eff}^2<0$
involves a competition between two effects: the
growth of the wavefunction, $\Psi\sim \exp(\omega t)$, and
the decrease of the coupling, $g_s=\exp(-Qt)$.
One way to quantify this is to think about the effect
of interaction terms in the Lagrangian \lscalar\ (with $Q\to -Q$).
The schematic structure of the full interacting Lagrangian
describing the field $T$ is\foot{For simplicity,
we include only the self-interactions of $T$;
similar comments apply to interactions of $T$ with other
fields.}
\eqn\lschem{
\CL=\half e^{2Qt} (a_2T^2+a_3T^3+a_4T^4+\cdots)~.
}
In terms of the wavefunctions $\Psi=T\exp(Qt)$,  
\lschem\ can be written as
\eqn\lspsi{
\CL=(b_2\Psi^2+b_3g_s\Psi^3+b_4g_s^2\Psi^4+\cdots)~.
}
Clearly, the effect of cubic and higher terms, relative to
the leading, quadratic, term in the action is governed
in the weak coupling region by the size of $T$. 
Thus, if $T$ diverges as $t\to+\infty$, the
late time solution is unstable against such perturbations, 
and one has to take into account tachyon condensation.

It is not difficult to see that if the mass of the
scalar field $T$ satisfies $m^2>0$, both solutions
for $T$ \basissol\ go to zero as $t\to+\infty$, \ie\
the late time behavior is stable. On
the other hand, for $m^2<0$, there is always a solution
for which $T$ grows exponentially at late times, and thus 
the late time behavior is modified by tachyon condensation.
It is reasonable to require that in a sensible vacuum
of string theory there should not be modes with $m^2<0$;
this is certainly the case in the type II examples
discussed in subsection {\it 3.1}.

For early times, all fields $T$ grow with time \basissol.
The question of (in)stability of the solution to generic
perturbations takes one outside the realm of the linear
dilaton part of spacetime, and we will not discuss it here.

\subsec{Infrared divergences at one loop}

In this subsection, we would like to show that modes
\basissol\ with $\vec k^2+m^2<Q^2$, which correspond to
exponentially growing wavefunctions \expwave\ at early and late times,
give rise to infrared divergences in one loop amplitudes
in string theory in asymptotically timelike linear dilaton
spacetimes. In fact, we have seen this already: in the
discussion of the bosonic string on such backgrounds we
pointed out that these modes contribute to the partition
sum the term \levmatch, which can be written as
$\exp(-\pi\tau_2(\vec k^2+m^2-Q^2))$. Thus, if
$m^2+\vec k^2<Q^2$, the integral
over $\tau_2$ diverges from the region near $\tau_2=\infty$,
which is an infrared divergence, usually associated with a tachyon.
A similar structure arises for the type II case \iipart.
As in the case of a constant dilaton, one can try to define
the divergent integral by analytic continuation \MarcusVS\
and interpret the resulting finite, complex result
as giving a decay rate \refs{\MarcusVS,\WeinbergVP}.

In the rest of this subsection we will discuss the one loop
amplitude in field theory. The two main motivations
for this discussion are the following. First,
in the next section we will study the case of
de Sitter space, which does not seem to have a satisfactory
string theory realization. There, we will have to use
field theoretic techniques. Second, we would like to
discuss the relation of the Euclidean worldsheet partition
sums \finbosz, \iipart\ to existing calculations in the
QFT literature which are done in Minkowski space, and appear
to be infrared finite.

As a warm-up exercise, consider the one-loop vacuum
diagram of a scalar field in Minkowski space (with
constant dilaton). In Schwinger parametrization, it
reads
\eqn\schwinger{Z_{\rm 1-loop}={\rm Tr}\int_0^\infty{ds\over
s}e^{-is(-\eta^{\mu\nu}\partial_\mu
\partial_\nu+m^2-i\epsilon)}~,
}
where the trace runs over an orthonormal basis of delta-function
normalizable modes, which we choose to be the following
eigenfunctions of the wave operator $-\eta^{\mu\nu}\partial_\mu
\partial_\nu+m^2$:
\eqn\basisnorm{
\{e^{iEt+i\vec k\cdot\vec x}|(E,\vec k)\in \Rop^{d+1}\}~.
}
Thus we find
\eqn\schwingerbis{
Z_{\rm 1-loop}=
\int dE\int d^{d}k
\int_0^\infty{ds\over s} e^{-is(-E^2+\vec
k^2+m^2-i\epsilon)}~.
}
Note that the $i\epsilon$ perscription makes the integral converge
at large $s$. Consider first the case $m^2\geq 0$. Then it is
convenient to Wick rotate (see \eg\ \PolchinskiRQ, p. 83)
\eqn\wick{
E \mapsto iE~;\ \ \ s\mapsto-is~,
}
so that \schwingerbis\ becomes
\eqn\schwingertris{
Z_{\rm 1-loop}=
i\int dE\int d^{d}k
\int_0^\infty{ds\over s} e^{-s(E^2+\vec
k^2+m^2)}~.
}
This expression clearly has no infrared (large $s$) divergences, like
the original Minkowski amplitude. Recall, also, that in string theory
the parameter $s$ in \schwingertris\ is proportional to (the imaginary
part of) the modulus of the worldsheet torus, $\tau_2$.

For the case $m^2<0$, it is not a priori clear that performing the
Wick rotation \wick\ is a sensible thing to do. In particular,
while \schwingerbis\ does not have large $s$ divergences,
the Wick rotated expression \schwingertris\ is IR divergent in
this case. Thus, one might be tempted to perform the continuation
\wick\ for all modes with positive $m^2$, and leave the contributions
of tachyonic modes in the original form \schwingerbis. One problem
with this is that one would then have to treat differently regions
in the spatial momentum integral where $\vec k^2+m^2$ is positive and
negative.

In string theory\foot{E.g. the critical bosonic string,
or type 0 string theory.} one is instructed to perform the
continuation to \schwingertris\ for all modes, including those
with $m^2<0$.
All but at most a few of the string modes have positive $m^2$,
and the treatment of the remaining ones is determined by the
requirement of modular invariance.

Another way to see this is the following. Suppose in string theory
one performed the continuation \wick\ for all the positive $m^2$
modes, and treated the negative $m^2$ modes differently. The
resulting partition sum $Z(\tau)$ (see \eg\ \iipart) would not be
modular invariant, and there would not be any justification to
integrate it over the modular domain which excludes the small $s$
region. One would thus integrate over $s$ between zero and infinity.
The resulting amplitude would have a divergence from {\it small} $s$,
which would be equivalent \KutasovSV\ to the original infrared divergence
due to the tachyon. Thus, in string theory, one cannot avoid the infrared
divergence associated with  \schwingertris\ with $m^2<0$. The best one
can do is to push it into a UV regime, but the two are related by the
standard UV/IR duality of perturbative string theory (worldsheet duality).

It is elementary to generalize the above discussion
to the timelike linear dilaton case. In the presence
of a dilaton $\Phi=-Qt$, the one loop partition
sum\foot{Other one loop amplitudes will behave in the
same way for the purposes of the present discussion.}
takes the form
\eqn\contr{ Z_{\rm 1-loop}={\rm
Tr}\int_0^\infty{ds\over s}e^{-is(-\eta^{\mu\nu}\partial_\mu
\partial_\nu+2Q\partial_0+m^2-i\epsilon)}~,
}
where the trace runs over a suitable basis of normalizable
functions. We choose the basis
\eqn\basis{\{e^{-Qt}e^{iEt+i\vec k\cdot\vec x}|(E,\vec k)\in
\Rop^{d+1}\}~,
}
such that \contr\ becomes
\eqn\contrbis{
\eqalign{Z_{\rm 1-loop}&=
\int dE\int d^{d}k
\int_0^\infty{ds\over s} e^{-is(-E^2+\vec k^2+m^2-Q^2-i\epsilon)}\cr
&=i\int dE\int d^{d}k
\int_0^\infty{ds\over s} e^{-s(E^2+\vec k^2+m^2-Q^2)}\cr
&=i\pi^{(d+1)/2}\int_0^\infty{ds\over s}s^{-(d+1)/2} e^{-s(m^2-Q^2)}~,
}}
where we again rotated the contour of integration over $s$ and $E$
as in \wick. We see that this indeed has a large $s$ divergence for
all $m^2<Q^2$. The treatment of these modes is again determined by
string theory, as above. This has been used implicitly in arriving at
\finbosz, \iipart.

\newsec{De Sitter space}

\lref\HullII{
C.~M.~Hull,
``de Sitter space in supergravity and M theory,''
JHEP {\bf 0111}, 012 (2001)
[hep-th/0109213].
}
\lref\BoussoFI{
R.~Bousso, O.~DeWolfe and R.~C.~Myers,
``Unbounded entropy in spacetimes with positive cosmological constant,''
hep-th/0205080.
}
We would now like to repeat the analysis of section~5 for de Sitter space
(see \eg\ \refs{\WittenKN,\MaldacenaMW,\BanksFE,\BoussoNF,\HullII,
\SpradlinPW,\SusskindRI,\BoussoFI}  and references therein for some
recent discussions of de Sitter space). In global coordinates, which cover
the whole de Sitter manifold, the metric reads
\eqn\globaldS{
ds^2=-d\tau^2+l^2{\rm cosh}^2(\tau/l)d\Omega_d^2~,
}
where $l$ is the de Sitter radius, $-\infty<\tau<\infty$,
and  $d\Omega_d^2$ is the line element on the unit $d$-sphere.
This metric solves
Einstein's equations with positive cosmological constant
\eqn\cosmo{
\Lambda={d(d-1)\over 2l^2}
}
and describes a space that contracts for early times $\tau$
and expands for late times. In that respect, it is
similar to the Einstein frame metric describing the solutions of
subsection~{\it 2.2} (see \einsteinmetric, \einsteinmetricbis\ for
the asymptotic behavior of this Einstein frame metric). We will
focus on the early time region
$\tau\rightarrow -\infty$. It will be convenient to describe this
region in planar coordinates, which cover half of de Sitter space:
\eqn\planarpast{
ds^2=-dt^2+e^{-2Ht}d\vec x^2~.
}
Here $-\infty<t<\infty$, $H=l^{-1}$ is the `Hubble constant', and
$\vec x$ are coordinates on a $d$-plane. Early times correspond to
$t\rightarrow-\infty$.

Translations of $t$ have to be
accompanied by dilations in order to preserve the metric. The
associated Killing vector $K$ is
\eqn\killing{
K=\partial_t+Hx^i\partial_i~,
}
which has norm squared
\eqn\normkilling{
(K,K)=-1+H^2e^{-2tH}\vec x^2~.
}
Thus we see that $K$ is timelike whenever $H^2e^{-2tH}\vec x^2<1$,
so that there is a cosmological event horizon of size
\eqn\horplan{
\vec x^2=H^{-2}e^{2tH}~.
}
In analogy with subsection~{\it 5.1}, we now study the classical
evolution of a scalar field with mass $m$
in the background \planarpast:
\eqn\mincoup{
S=\half\int dt\int d^dx\, e^{-dHt}\, (\dot T^2-e^{2Ht}(\nabla T)^2
-m^2 T^2)~,
}
with equation of motion
\eqn\eomds{
\ddot T-dH\dot T-e^{2Ht}\nabla^2 T+m^2T=0~.
}
We want to study this scalar field for early times $t$, for which
the $e^{2Ht}\nabla^2 T$ term in \eomds\ is negligible.\foot{We
will comment on what this means shortly.} Then the solutions are
\eqn\sol{
T_\pm\sim e^{h_\pm t}e^{i\vec k\cdot\vec x}\ ,\ \ t\rightarrow-\infty~,
}
where
\eqn\hplusminus{
h_\pm={dH\over2}\pm\sqrt{({dH\over2})^2-m^2}~.
}
For $0<m^2<({dH\over2})^2$, both solutions grow exponentially with $t$
for early times. One can again rescale the field $T$ in \mincoup,
$T=\Psi\exp(dHt/2)$, and find that the action \mincoup\ (with the
second term and a total derivative neglected) takes the form
\eqn\mincoup{
S=\half\int dt\int d^dx\,(\dot \Psi^2-m_{\rm eff}^2 \Psi^2)~,
}
with $m_{\rm eff}^2=m^2-(dH/2)^2$. Hence, growing modes exist when
\eqn\growham{
m^2<({dH\over 2})^2~.
}
In the analysis leading to \sol, we assumed that $t$ was
sufficiently large and negative for the $e^{2Ht}\nabla^2 T$
term in \eomds\ to be negligible compared to the other two
terms in the action. This is the case if
\eqn\ineq{
e^{2Ht}\vec k^2<<m^2,\;({dH\over 2})^2~.
}
Using \horplan, we see that our approximation is valid only for
modes with wavelength large compared to the horizon radius (up to a
$d$-dependent factor). This should be contrasted with discussions
like that of \Abbott, where fluctuations inside the event horizon
are discussed. 

As in subsection~{\it 5.2}, the growing modes \growham\
lead to infrared divergences in one-loop amplitudes.
The contribution of the scalar field \mincoup\ to the one-loop
vacuum amplitude is
\eqn\contrds{
Z_{\rm 1-loop}={\rm Tr}\int_0^\infty{ds\over s}e^{-is(-\partial_t^2
+dH\partial_t+e^{2Ht}\nabla^2+m^2-i\epsilon)}~,
}
where the trace runs over a suitable basis of normalizable
functions. We choose these functions to be
eigenfunctions of the differential operator in the exponent in \contrds,
and label them by their behavior for early $t$, which is given by
\eqn\basis{
\{e^{{dHt\over 2}}e^{iEt+i\vec k\cdot\vec x}|
(E,\vec k)\in \Rop^{d+1}\}~.
}
We evaluate the trace in the early time region,
so that the $e^{2Ht}\nabla^2$ term in
\contrds\ is negligible. Then \contrds\ becomes
\eqn\contrdsbis{
Z_{\rm 1-loop}=
\int dE\int d^{d}k
\int_0^\infty{ds\over s} e^{-is(-E^2+m^2-({dH\over2})^2-i\epsilon)}~,
}
where the integral over $\vec k$ has a large $|\vec k|$ cutoff
determined by the condition that the $e^{2Ht}\nabla^2$ term in
\contrds\ should be negligible. The result of the
integral over $\vec k$ is a factor converting the coordinate
volume of space (which multiplies the right hand side of
\contrdsbis, although we have suppressed it in the formulae) into the
physical volume of space (\ie, the volume measured with the metric
\planarpast).
Omitting this volume factor, we obtain by Wick rotation (as in
subsection~{\it 5.2})
\eqn\eval{
Z_{\rm 1-loop}=
i\int dE\int_0^\infty{ds\over s} e^{-s(E^2+m^2-({dH\over2})^2)}~.
}
We see that \eval\ has large $s$ divergences for all masses with
$m^2<({dH\over2})^2$. This is precisely the range of masses 
corresponding to growing modes \growham.

\newsec{Summary and discussion}


The main purpose of this paper was to study string propagation
in the time-dependent spacetime \bhuv, \uvid, which includes a
cosmological singularity, in the vicinity of which spacetime has
the form $\IR^{1,1}/\IZ$. In such spacetimes there are many
qualitative and quantitative issues that are not well understood
in the framework of QFT in curved spacetime. For example, it is
not clear whether one should only include the part of the spacetime
which looks like a circle expanding from zero size at the singularity
to a finite size at late times, or also the pre-big bang region,
as well as the other regions in figures 1,2. There is a question as
to the nature of observables, and in particular the continuation of
wavefunctions through cosmological singularities.
Ultimately, one would like to compute S-matrix elements in such
spacetimes, and study the effects of the cosmological singularities
on them.

We found that many of these and other questions can be addressed
by realizing the cosmological spacetime \bhuv, \uvid\ as a coset
CFT of the form $SL(2,\IR)/(U(1)\times \IZ)$. Our main results are
the following:

The coset CFT describes a spacetime consisting of six regions (see
figures 1,2), four of which include asymptotic early and late time
regimes (two of each kind). The observables correspond to scattering
states prepared at early times, and their S-matrix elements to
evolve to some particular states at late times.

The asymptotic past consists of two disconnected regions, regions
$I$ and $VI$ in figures 1,2. Correspondingly, the Hilbert space
of asymptotic past states naturally takes the form of a direct product
of the two Hilbert spaces corresponding to the two regions.
Similarly, the future Hilbert space is a direct product of the
Hilbert spaces corresponding to regions $II$ and $V$. In the asymptotic
past and future the solution approaches a timelike linear dilaton
one. The string coupling goes to zero, which makes it relatively
easy to construct the incoming and outgoing Hilbert spaces.

Embedding the cosmological spacetime in $SL(2,\IR)$ allows one
to continue wavefunctions through singularities. We used this
continuation to compute the Bogolubov coefficients relating
particles in the initial and final Hilbert spaces. We found
that in order to obtain unitary evolution in the cosmological
spacetime, one has to include both of the incoming and both
of the outgoing regions. We also showed that the incoming vacuum
evolves to a state with particles in the far future.

We discussed the existence of growing modes in asymptotically timelike
linear dilaton solutions. We showed that scalar fields with mass
$m^2<Q^2$ have exponentially growing or decaying wavefunctions (for
low enough momentum), and lead to infrared divergences in one loop amplitudes.

We briefly discussed the relation of our results to gravity in
de Sitter space. In the early and late time regimes, the qualitative
structure of solutions to the wave equation in timelike linear dilaton
and global de Sitter spacetime is similar, and thus one can use some
of the arguments made in the timelike linear dilaton case. One finds
that fields with mass smaller than the Hubble mass lead to  modes
with growing wavefunction in de Sitter space, which are analogous to
those with $m^2<Q^2$ in timelike linear dilaton backgrounds.
In particular, they lead to infrared divergences in one loop diagrams.

An issue that has
received some attention recently is the question of
holography in de Sitter space, and in particular the proposal
\StromingerPN\ that gravity in de Sitter space is equivalent to
a Euclidean CFT living on the spacelike boundary at early and late
times. The analogous statement in asymptotically timelike linear dilaton
spacetimes would be that the theory is equivalent to a Euclidean
theory living on the spacelike boundary at $|t|\to\infty$.

From our discussion it seems that the situation in both of these cases
is expected to be more analogous to gravity in asymptotically flat
spacetime than to that in anti-de Sitter or spacelike linear dilaton
vacua. In the latter case, there is a class of observables that
correspond to non-normalizable wavefunctions supported at the boundary,
and one can study the path integral of gravity as a function of
these ``boundary conditions''. In flat spacetime, one instead studies
(delta function) normalizable wavefunctions that correspond to scattering
states and their S-matrix.

In de Sitter and timelike linear dilaton spacetimes, the analogues of
the non-normalizable wavefunctions of AdS and spacelike linear dilaton
solutions seem to be the growing modes discussed in sections 5,6.
The ``good observables'' are in fact
analogues of the scattering states in flat spacetime. Therefore,
it is not completely clear that a picture in terms of a Euclidean theory
living on the spacelike boundary will be more useful in this case
than an analogous picture in flat spacetime. A better understanding of
holography in these spacetimes would be interesting.

Many other issues deserve further study. We described
in section 2 a large class of solutions, the generalized Kasner solutions
\highphi, \rrii. If the parameters $a_i$ \aaii\ are small, one can think
of the resulting solutions as ``small perturbations'' of the generalized
Milne solution that we discussed in sections 3,4 using coset CFT techniques.
These perturbations correspond to zero momentum modes of the graviton and
dilaton, which are massless but become effectively tachyonic in the
timelike linear dilaton spacetime, \meffec. The Kasner solutions are
in fact an example of how such modes can significantly influence the evolution,
even if they are very small at early times. The 
uncompactified generalized Milne solution
\bhmet\ is non-singular at $t=0$ (as discussed in subsection {\it 2.3}),
while arbitrarily small perturbations of it at early times, corresponding to
turning on $a_2, a_3,\ldots$ lead to a singularity at finite $t$.
It would be interesting to understand whether all the generalized Kasner
solutions can be understood by thinking of the spacetime as a perturbed
coset model. In particular, it would be interesting to understand
whether one still has to include the different regions that were seen
to play a role in the generalized Milne case, and if so, how to continue
wavefunctions through the Kasner singularities.

Another general question involves the supercritical type II solutions
corresponding to asymptotically timelike linear dilaton spacetimes
at $t\to+\infty$, discussed in \refs{\AntoniadisAA,\AntoniadisVI} and in
subsection {\it 3.1}. Since these solutions do not contain fields with
$m^2<0$, it seems that their late time behavior might be stable. One
can then ask what the status of these solutions is within M-theory. They are 
obviously higher than eleven dimensional, and at least naively seem to have 
more degrees of freedom than other known vacua of M-theory. It is not clear 
how they are related to the standard eleven dimensional descriptions of M-theory.

\bigskip
\noindent{\bf Acknowledgements:}
We would like to thank I.~Bars, S.~Elitzur, A.~Giveon, S.~Hollands,
P.~Kraus, F.~Larsen, E.~Martinec, R.~Myers, E.~Rabinovici, A.~Schwimmer,
N.~Seiberg and R.~Wald for discussions. This work is supported in part
by DOE grant DE-FG02-90ER40560 and by NSF grant PHY-9901194. G.R.
would like to thank the theory division at LBNL and UC Berkeley
for hospitality while this work was being completed.

\listrefs
\end